\documentclass[twocolumn,preprintnumbers,floats,prd,amssymb,floatfix,nofootinbib,balancelastpage,superscriptaddress,amsmath]{revtex4-1}

\pdfoutput=1
\usepackage[utf8]{inputenc}

\usepackage{amssymb}
\usepackage{amsmath}
\usepackage{amsfonts}
\usepackage{graphicx}
\usepackage{xcolor}
\usepackage[normalem]{ulem}
\usepackage{xspace}

\usepackage[section]{placeins}
\usepackage{afterpage}

\usepackage{upgreek}
\usepackage{braket}
\usepackage{float}
\usepackage{slashed}
\usepackage{multirow,rotating}

\def\gsim{\raise0.3ex\hbox{$\;>$\kern-0.75em\raise-1.1ex\hbox{$\sim\;$}}}
\def\lsim{\raise0.3ex\hbox{$\;<$\kern-0.75em\raise-1.1ex\hbox{$\sim\;$}}}

\newcommand{\lam}{\lambda}

\newcommand{\AddrBonn}{%
Bethe Center for Theoretical Physics \& Physikalisches Institut der 
Universit\"at Bonn,\\ Nu{\ss}allee 12, 
 53115 Bonn, Germany}

\newcommand{\AddrAPCTP}{%
	Asia Pacific Center for Theoretical Physics (APCTP) 
	- Headquarters San 31,\\ Hyoja-dong, Nam-gu, Pohang 790-784, Korea}

\newcommand{\AddrNTHU}{%
	Department of Physics, National Tsing Hua University, Hsinchu 300, Taiwan
	}

\def\gsim{\raise0.3ex\hbox{$\;>$\kern-0.75em\raise-1.1ex\hbox{$\sim\;$}}}
\def\lsim{\raise0.3ex\hbox{$\;<$\kern-0.75em\raise-1.1ex\hbox{$\sim\;$}}}

\begin{document}

\preprint{APCTP Pre2020-019}
\preprint{BONN-TH-2020-07}

\title{R-parity Violation and Light Neutralinos at ANUBIS and MAPP}

\author{Herbert K. Dreiner}
\email{dreiner@uni-bonn.de}
\affiliation{\AddrBonn}

\author{Julian Y. G\"unther}
\email{s6juguen@uni-bonn.de}
\affiliation{\AddrBonn}

\author{Zeren Simon Wang}
\email{wzs@mx.nthu.edu.tw}
\affiliation{\AddrNTHU}
\affiliation{\AddrAPCTP}

%%%%%%%%%%%%%%%%%%%%%%%%%%%%%%%%%%%%%%%%%%%%%%%%%%%%%%%%%%%%%%%%%%%%%%
\begin{abstract}
	In  R-parity-violating supersymmetry the lightest neutralino can be very light, even massless. For masses in the range $500$ MeV$\lsim 
	m_{\tilde\chi^0_1}\lsim 4.5$ GeV the neutralino can be produced in hadron collisions from rare meson decays via an R-parity violating coupling, 
	and subsequently decay to a lighter meson and a charged lepton. Due to the small neutralino mass and for small R-parity violating coupling the 
	lightest neutralino is long-lived, leading to displaced vertices at fixed-target and collider experiments. In this work, we study such signatures at the 
	proposed experiments  ANUBIS and MoEDAL-MAPP at the LHC. We also compare their sensitivity reach in these scenarios with that of other 
	present and proposed experiments at the LHC such as ATLAS,  CODEX-b, and  MATHUSLA. We find that  ANUBIS
	and  MAPP can show complementary or superior sensitivity.
\end{abstract}
%%%%%%%%%%%%%%%%%%%%%%%%%%%%%%%%%%%%%%%%%%%%%%%%%%%%%%%%%%%%%%%%%%%%%%
\keywords{RPV-MSSM, neutralinos, LHC, LLP}

%\arxivnumber{}
%\pacs{14.60.Pq, 12.60.Jv, 14.80.Cp}

\vskip10mm

\maketitle
\flushbottom
%%%%%%%%%%%%%%%%%%%%%%%%%%%%%%%%%%%%%%%%%%%%%%%%%%%%%%%%%%%%%%%%%%%%%%
%\tableofcontents
%
%%%%%%%%%%%%%%%%%%%%%%%%%%%%%%%%%%%%%%%%%%%%%%%%%%%%%%%%%%%%%%%%%%%%%%
%%%%%%%%%%%%%%%%%%%%%%%%%%%%%%%%%%%%%%%%%%%%%%%%%%%%%%%%%%%%%%%%%%%%%%
%\tableofcontents
%
%%%%%%%%%%%%%%%%%%%%%%%%%%%%%%%%%%%%%%%%%%%%%%%%%%%%%%%%%%%%%%%%%%%%%%
\section{Introduction}
\label{sect:intro}

Recently there has been an increased interest in long-lived particles (LLPs). Such particles are defined at colliders to have detached 
vertices (DVs). Instead of promptly decaying after production, they travel for a macroscopic distance before decaying within the detector, or in
nearby additional detectors. Such a long lifetime can arise for different reasons such as small mass splitting, feeble couplings, or a heavy mediator. 
While LLPs exist already in the Standard Model (SM), such as the long-lived hadron $K_L$, they are also frequently predicted in a variety of BSM 
models motivated by either dark matter or the non-vanishing neutrino masses. For example, portal-physics models connecting the SM and dark 
sectors may lead to such LLPs which have a tiny coupling with the SM particles. These models may include dark photons (vector portal), a light 
scalar (Higgs portal), axion-like particles (pseudoscalar portal), or heavy neutral leptons (fermion portal). Moreover, other theoretical scenarios such 
as quirky models and split supersymmetry (SUSY) models also predict LLPs. For recent reviews of LLP models and studies, see 
Refs.~\cite{Alimena:2019zri,Lee:2018pag,Curtin:2018mvb}.

We are here interested in supersymmetric models with light neutralinos. Searches for promptly decaying heavy supersymmetric fields have been 
unsuccessful so far. Lower limits on the masses of squarks and gluinos have been placed at the order of TeV in various SUSY models. However, this is 
not the case for the lightest neutralino, $\tilde{\chi}_1^0$. It was noticed some time ago \cite{Choudhury:1995pj,Choudhury:1999tn}, that if we drop the 
GUT (grand unified theory) motivated relation of the gaugino masses $M_1 = \frac{5}{3}\tan^2{\theta_W}M_2$ and drop the dark matter constraint on the 
lightest neutralino \cite{Belanger:2002nr,Hooper:2002nq,Bottino:2002ry,Belanger:2003wb,Vasquez:2010ru,Calibbi:2013poa}, then the neutralino mass can be 
below a GeV and even massless \cite{Gogoladze:2002xp,Dreiner:2009ic}. Such a light neutralino is consistent with stellar cooling, of supernov{\ae} 
\cite{Grifols:1988fw,Ellis:1988aa,Lau:1993vf,Dreiner:2003wh}, and of white dwarfs \cite{Dreiner:2013tja}, as well as with cosmology 
\cite{Profumo:2008yg,Dreiner:2011fp}. Such light neutralinos, if stable, result in a relic energy density overclosing the Universe \cite{Bechtle:2015nua}. 
Thus they must decay. In R-parity violating supersymmetry (RPV-SUSY) models, see Refs.~\cite{Dreiner:1997uz,Barbier:2004ez,Mohapatra:2015fua} 
for reviews, the lightest neutralino decays via the RPV couplings. When both the mass of the lightest neutralino and the RPV couplings are sufficiently 
small, the lightest neutralino is long-lived.

Searches for light long-lived neutralinos have been studied in various experimental setups. These include existing fixed-target experiments \cite{Choudhury:1999tn,Dedes:2001zia,Dreiner:2002xg,Dreiner:2009er}, a proposed new fixed-target experiment: \texttt{SHiP} at CERN
\cite{Alekhin:2015byh,Gorbunov:2015mba,deVries:2015mfw}, the \texttt{ATLAS} experiment \cite{deVries:2015mfw} and a variety of proposed dedicated 
experiments at the LHC: \texttt{CODEX-b} \cite{Gligorov:2017nwh,Aielli:2019ivi}, \texttt{FASER} \cite{Feng:2017uoz,Ariga:2018uku,Ariga:2019ufm}, 
\texttt{MATHUSLA} \cite{Chou:2016lxi,Curtin:2018mvb}, and \texttt{AL3X} \cite{Gligorov:2018vkc}) \cite{Helo:2018qej,Dercks:2018eua,Dercks:2018wum}, and 
future $Z-$factories \cite{Wang:2019orr,Wang:2019xvx}.\footnote{Both \texttt{ATLAS} and  \texttt{CMS} have searched for heavier long-lived neutralinos. 
The hadronic decays are then to jets instead of light mesons, as we consider here. See for example Refs.~\cite{Sirunyan:2019gut,Aad:2019xav} and references 
therein.} In this work, we consider two relatively new proposals of dedicated experiments for searching for neutral LLPs at the LHC, namely \texttt{ANUBIS} 
(``An Underground Belayed In-Shaft experiment'') \cite{Bauer:2019vqk,Hirsch:2020klk} and \texttt{MAPP} (``\texttt{MoEDAL} Apparatus for the detection of 
Penetrating Particles'') \cite{Staelens:2019gzt}. \texttt{ANUBIS} is to consist of a cylindrical detector installed inside one of the service shafts 
above either the \texttt{ATLAS} or \texttt{CMS} interaction point (IP), with an expected integrated luminosity of 3 ab$^{-1}$. \texttt{MAPP} is planned with two 
phases and to be installed inside the UGCI gallery near the interaction point 8 (IP8) of the LHC, where the experiment \texttt{LHCb} is located. \texttt{MAPP1}
 and \texttt{MAPP2} are projected to have an integrated luminosity of 30 and 300 fb$^{-1}$, respectively. The details of these three experiments are discussed 
in Sec.~\ref{sect:exp}.

In the existing literature on the search for long-lived light neutralinos at various experiments, two types of production mechanisms have been considered. The first is rare $Z-$boson decays into a pair of the lightest neutralinos via the small Higgsino component \cite{Bartl:1988cn,Helo:2018qej,Dercks:2018wum,Wang:2019orr,Wang:2019xvx}, and the second is rare meson decays into a single 
neutralino plus a neutral or charged lepton via an RPV coupling \cite{Choudhury:1999tn,Dedes:2001zia,Dreiner:2002xg,Dreiner:2009er,deVries:2015mfw,Dercks:2018eua,Dercks:2018wum}. 
In this work, we focus on the long-lived light neutralinos in RPV-SUSY, produced from a rare 
charm or bottom meson decay. The neutralino decays via an RPV coupling, again to a meson and a lepton. Taking one benchmark scenario for charmed and 
bottomed mesons, respectively, we compare the sensitivity reach of \texttt{ANUBIS} and \texttt{MAPP} experiments with other present and proposed experiments 
at the LHC.

This paper is organized as follows. In Sec.~\ref{sect:theory} we introduce the model basics of RPV-SUSY and the lightest neutralino. In Sec.~\ref{sect:exp}, we 
introduce the detector setups of \texttt{ANUBIS} and \texttt{MAPP}, and explain the simulation procedure and signal estimation. The numerical results for two 
benchmark scenarios are presented in Sec.~\ref{sect:results}. In Sec.~\ref{sect:conclusion} we summarize our findings and provide an outlook.

\section{Model basics of RPV-SUSY, the production and decay of the $\tilde{\chi}_1^0$}
\label{sect:theory}

Here, we introduce the RPV-SUSY model, and discuss the production and decay of the lightest neutralinos via 
RPV couplings. In RPV-SUSY, the MSSM superpotential is extended by the following renormalizable terms:
\begin{eqnarray}
W_{\text{RPV}} =&& \kappa_i L_i H_u + \frac{1}{2}\lambda_{ijk} L_i L_j E^c_k + \lambda'_{ijk} L_i Q_j D^c_k \nonumber\\
&&+ \frac{1}{2}\lambda^{''}_{ijk} U^c_i D^c_j D^c_k\,.
\label{eqn:RPVsuperpotential}
\end{eqnarray}
Here we use the notation as in Ref.~\cite{Allanach:2003eb}. In particular the $\lam,\,\lam',\,\lam''$ are dimensionless Yukawa couplings, 
and $i,j,k\in\{1,2,3\}$ are generation indices. The first three sets of terms violate lepton number and the last violates baryon number. In 
order to avoid rapid proton decay we consider an additional baryon triality, B$_3$, symmetry imposed \cite{Ibanez:1991pr,Dreiner:2012ae}, 
which allows only the lepton-number violating operators and is discrete gauge anomaly-free. For this work, we choose to consider only the 
$LQ\bar{D}$ operators. The lightest supersymmetric particle (LSP) is then no longer stable and decays into SM particles. In this study, we assume that the lightest neutralino 
is the LSP, which it need not be \cite{Dreiner:2008ca,Dercks:2017lfq}.

Following Refs.~\cite{deVries:2015mfw,Dercks:2018eua}, we investigate 2 benchmark scenarios, where the $\tilde{\chi}_1^0$ LSP's are 
singly produced from either a charm or a bottom meson's rare decay, and then decay to a lighter meson with a displaced vertex to be 
reconstructed inside a detector. Such light GeV-scale, or lighter, neutralinos are necessarily binolike to avoid existing 
bounds~\cite{Gogoladze:2002xp,Dreiner:2009ic}. We perform the computation of the decay widths of the heavy mesons into the lightest 
neutralino and of the neutralinos into a lighter meson plus a neutral or charged lepton, with the analytic formulas given in 
Refs.~\cite{Choudhury:1999tn,Dedes:2001zia, Dreiner:2009ic, deVries:2015mfw}. In each of these benchmark scenarios, two RPV 
couplings are assumed to be nonzero, responsible for the production and decay of $\tilde{\chi}_1^0$, respectively. We work directly at the 
low-energy scale, disregarding the possibility of multiple RPV couplings generated as a result of the renormalization group equations~\cite{Allanach:1999mh}.

The RPV couplings are in general constrained by various experimentally measured observables. Such bounds usually depend on the relevant 
scalar fermion particles. See Refs.~\cite{Barbier:2004ez,Allanach:1999ic,Barger:1989rk,Bhattacharyya:1997vv,Kao:2009fg,Domingo:2018qfg} 
for reviews. Below we summarize the current bounds on both the single RPV couplings and coupling products that are relevant to the benchmark 
scenarios we study, extracted from Refs.~\cite{Allanach:1999ic,Kao:2009fg,Domingo:2018qfg,Bansal:2019zak}.

The current single couplings bounds are \cite{Kao:2009fg,Bansal:2019zak}:
\begin{eqnarray}
&\lambda'_{112}<0.030+0.16 \,\frac{m_{\tilde{d}_R}}{1 \text{ TeV}}\,,\label{eqn:singleRPVbounds1}\\
&\lambda'_{122}<2 \,\frac{m_{\tilde{s}_R}}{1 \text{ TeV}}\,,\label{eqn:singleRPVbounds2} \\
&\lambda'_{131} < 0.19\, \frac{m_{\tilde{t}_L}}{1\text{ TeV}}\,,\label{eqn:singleRPVbounds3}
\end{eqnarray}
and coupling products bounds are \cite{Allanach:1999ic,Domingo:2018qfg}:
\begin{eqnarray}
&\sqrt{\lambda'_{122}\lambda'_{112}} < 4.7\times 10^{-2} \,\frac{m_{\tilde{s}_R}}{1 \text{ TeV}}\,,\label{eqn:RPVproductbounds1}\\
&\sqrt{\lambda'_{131}\lambda'_{112}} < 3.0\times 10^{-3} \,\frac{m_{\tilde{e}_L}}{1 \text{ TeV}}\,.\label{eqn:RPVproductbounds2}
\end{eqnarray}
These bounds on the RPV couplings stem from different phenomenological origins, including meson decays and oscillations, atomic parity 
violation, as well as LHC Drell-Yan data and electroweak precision measurements from LEP and SLC. Consequently they depend on the 
masses of different sfermions. In this work, we assume for simplicity degenerate sfermion masses, and will compare the sensitivity of 
\texttt{ANUBIS} and \texttt{MAPP} in the parameter space of the RPV-SUSY with the current experimental bounds.
All the above bounds derive from R-parity conserving reactions, involving two insertions of an R-parity violating operator and one supersymmetric scalar propagator. Thus in each case the amplitude is proportional to $\lam^{\prime2}/{\tilde m}^2$. The meson decays we consider below violate R-parity, as the initial state is R-parity even and the final state with one neutralino is R-parity odd. Thus the amplitudes are proportional to $\lam'/{\tilde m}^2$, manifesting a different scaling behaviour.

Before we move to the next section, we briefly summarize the current lower bounds on sfermion masses derived from direct SUSY searches at the LHC, for the $LQ\bar{D}$ couplings considered in this work \cite{Dercks:2017lfq}. We find that while there is presently no relevant constraint on sneutrino, selectron, and sbottom masses, lower limits on the masses of first- and second-generation squarks and stops exist for $\lambda'_{112}$ and $\lambda'_{122}$ (though not for $\lambda'_{131}$).
Assuming $m_{\tilde{\chi}_1^0}=0.5  \, m_{\tilde{q}}$ or $m_{\tilde{\chi}_1^0}=0.9  \, m_{\tilde{q}}$, the bounds on squark masses for $\lambda'_{112}$ and $\lambda'_{122}$ are 1160 and 1315~GeV, respectively \cite{ATLAS:2015gma}. Further, the present bounds on $m_{\tilde{t}}$ are 890~GeV for $\lambda'_{1bc}$ with $b,c \in \{1,2\}$, assuming $m_{{\chi}_1^\pm}=100$ GeV \cite{Khachatryan:2016ycy}.
Since these bounds are in general around 1 TeV, we will focus on two benchmark values of the degenerate sfermion masses, 1\,TeV and 5\,TeV, when we compare the sensitivity reach of \texttt{ANUBIS} and \texttt{MAPP} to the existing limits in Sec.~\ref{sect:results}.

\section{Experimental setups and simulation procedure} 
\label{sect:exp}

In this section we introduce the detector setups of the proposed experiments \texttt{ANUBIS} and \texttt{MAPP},
explain the simulation procedure, and discuss the estimate of signal-event numbers.

At both experiments, there are various potential background sources such as long-lived SM hadrons decays and 
cosmic rays. Such background events can be effectively reduced to the negligible level by \textit{e.g.} 
charged-particle vetos and directional cuts, as discussed in Refs.~\cite{Bauer:2019vqk,Staelens:2019gzt}.
Accordingly we assume 0 background events for the sensitivity study in this paper. Furthermore, since the detailed 
detector information of these two experiments are not yet available, for simplicity we assume 100\% detector 
efficiencies here.
 
\subsection{ANUBIS and MAPP}

\begin{figure}[t]
	\includegraphics[width=0.5\textwidth]{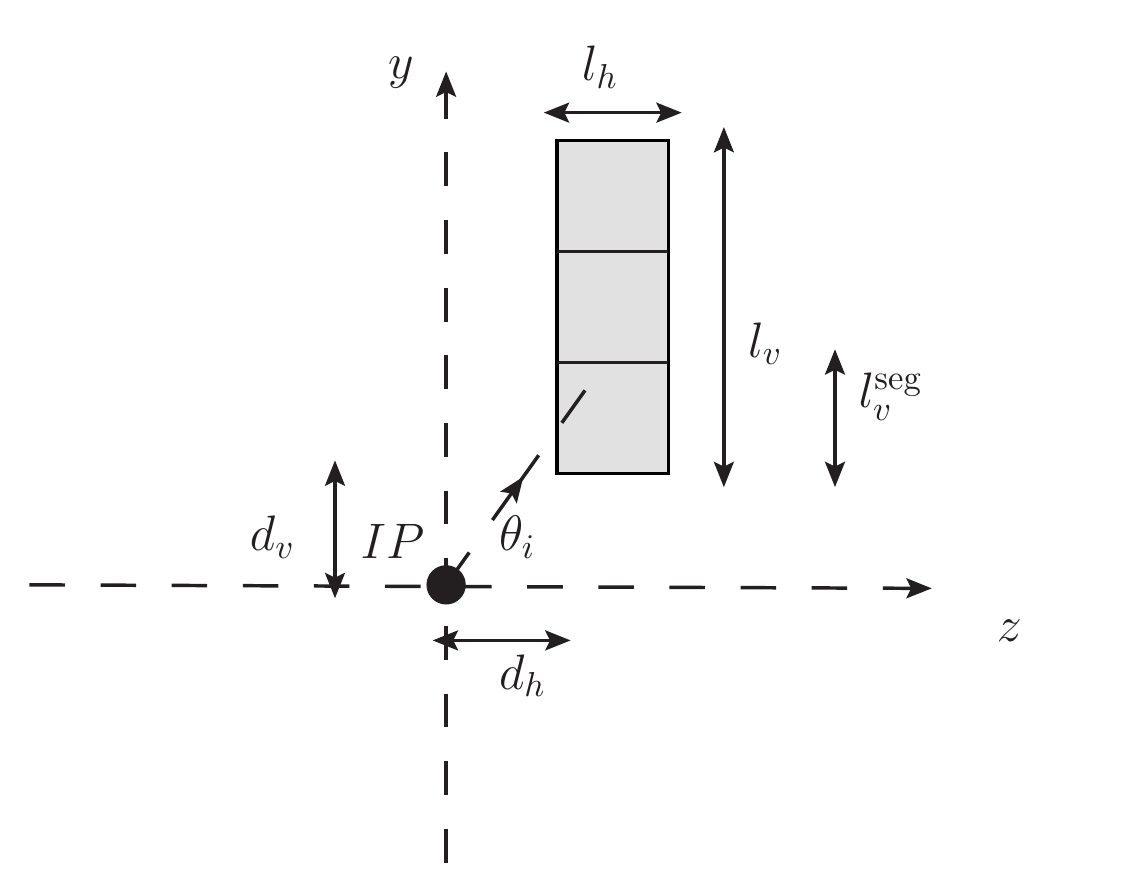}
	\includegraphics[width=0.5\textwidth]{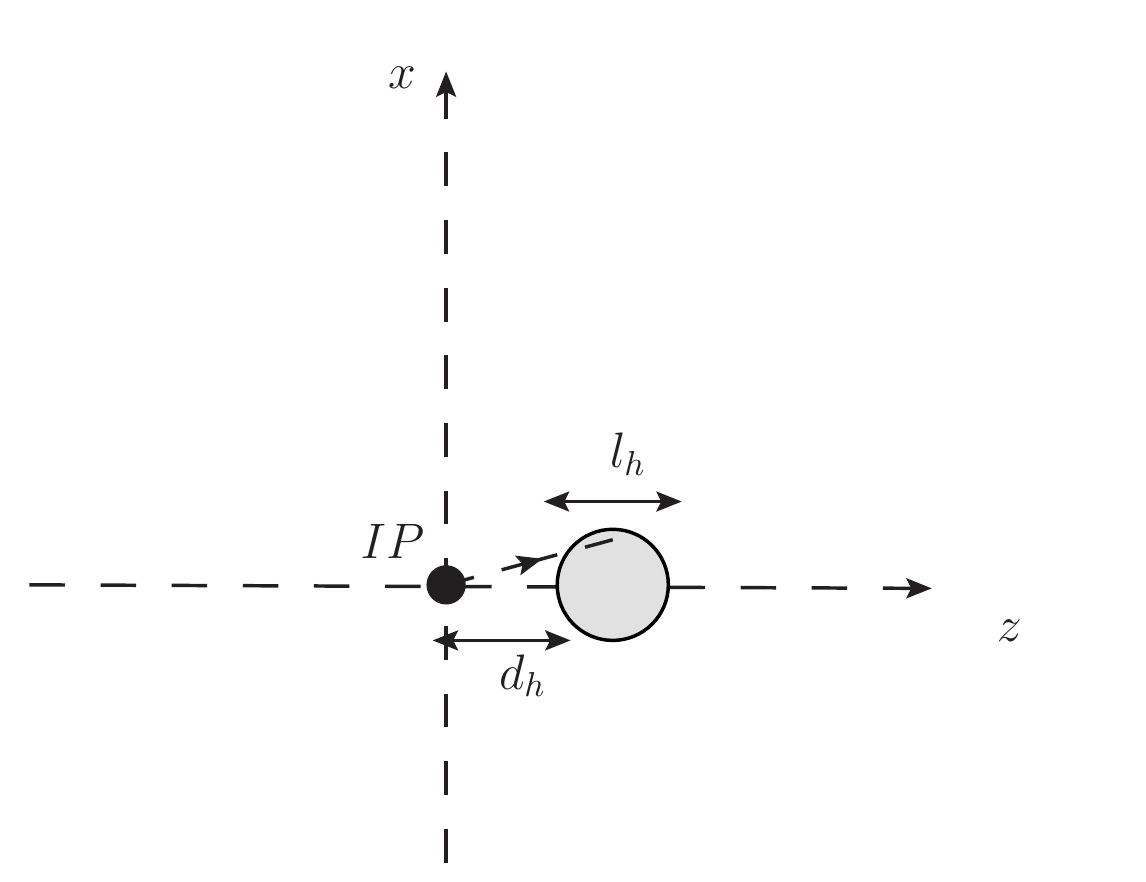}
	\caption{	The profile sketches of the ANUBIS detector in the $y-z$ (looking from the side) and the $x-z$ (looking from the top down) planes, respectively, extracted from Ref.~\cite{Hirsch:2020klk}. A sample LLP-event with polar angle $\theta_i$ is included in the sketches.
			}
	\label{fig:ANUBISsketch}
\end{figure}

\texttt{ANUBIS} is proposed as a cylindrical detector making 
use of one of the installation shafts at the \texttt{ATLAS} or \texttt{CMS} IP.  Sketches of the experiment, 
reproduced from  Ref.~\cite{Hirsch:2020klk}, are presented in Fig.~\ref{fig:ANUBISsketch} from two perspectives, where a 
sample LLP with polar angle $\theta_i$ is labeled with a dashed arrow. \texttt{ANUBIS} has a height, $l_v$, of 56 m and a 
diameter $l_h$ of 18 m, so that the fiducial region consists of approximately $\sim 14,\!250\,\text{m}^3$. It has a horizontal 
(vertical) distance $d_h$ ($d_v$) of 5 (24) m from the IP. Four tracking stations are planned to be installed in parallel, with 
intervals of 18.5\,m.

Compared to the other proposed dedicated far-detector experiments at the LHC, \texttt{ANUBIS} has several advantages. 
First, with its location inside one of the service shafts above the \texttt{ATLAS}/\texttt{CMS} IP, it can be particularly sensitive 
to LLPs traveling at a larger polar angle. It can be integrated directly with the \texttt{ATLAS}/\texttt{CMS} experiment, 
extending the sensitivity of these currently running experiments. A total integrated luminosity as large as 3~ab$^{-1}$ at the 
HL-LHC is expected for the \texttt{ANUBIS} experiment.

\texttt{MAPP}  is located in the UGCI gallery at the IP8 at the LHC, close to the \texttt{MoEDAL} detector (``Monopole and 
Exotics Detector At the LHC''). The first phase of the experiment known as \texttt{MAPP1} is planned to be in operation during 
the LHC RUN-3 with an integrated luminosity of $30$ $\mathrm{fb}^{-1}$. \texttt{MAPP1} consists of two sub-detectors: 
\texttt{MAPP-mCP} to detect minimally charged particles and \texttt{MAPP-LLP} to search for neutral LLPs. We consider the 
latter with an approximate fiducial volume of $\sim$ 130 m$^3$. 

The detector can be placed at multiple positions in the UGCI gallery with an angular range of $5^\circ$ to $25^\circ$. 
Depending on the angle, the detector is shielded by 25 to 55 m of rock. We consider the position at $5^\circ$ with a 
distance of 55 m. Additionally, the detector is shielded by 100 m of rock above it, so that we assume no background for 
the \texttt{MAPP} experiment.

\begin{figure}
	\includegraphics[width=0.45\textwidth]{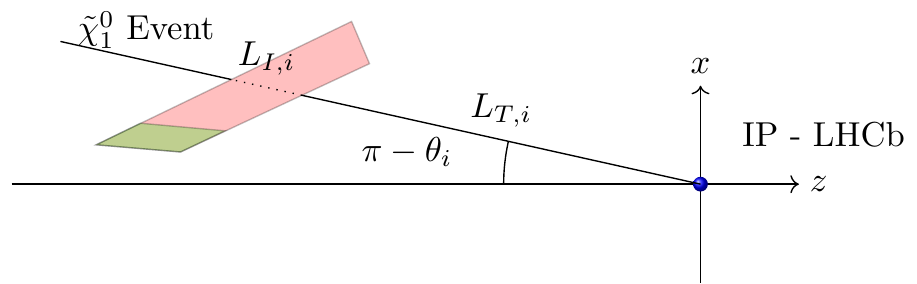}
	\includegraphics[width=0.45\textwidth]{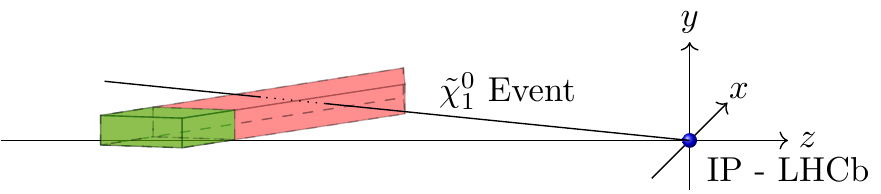}
	\caption{Sketches of \texttt{MAPP1} (green) and \texttt{MAPP2} (red+green) in the $x$-$z$ plane, \textit{i.e.} viewed 
	from above, and in three dimensions, with the $y$-axis pointing vertically upwards. A sample event is shown hitting 
	\texttt{MAPP2} but not \texttt{MAPP1}.}
	\label{fig:MAPPsketch}
\end{figure}

After RUN-3 an upgrade of the detector known as \texttt{MAPP2} is planned, in which the fiducial region is extended to cover almost the whole gallery with a volume of $\sim$ 430 m$^3$.
The sketches of both detectors are given in Fig.~\ref{fig:MAPPsketch}, where the \texttt{MAPP1} detector is shown in green only while  \texttt{MAPP2} will occupy 
both the green and red regions. Similar to Fig.~\ref{fig:ANUBISsketch}, a sample event with polar angle $\theta_i$ is illustrated 
in the figures. 

\subsection{Simulation and Signal Event Estimate}

We proceed to describe the simulation procedure and the estimate of the number of signal events in each of these experiments.

Since we consider only the production of the lightest neutralinos from rare decays of charm and bottom mesons, we can express 
the total number of produced $\tilde{\chi}_1^0$'s in terms of the total number of produced mesons $N_M$, their lifetime $\tau_M$, 
and the meson partial  decay width into $\tilde{\chi}_1^0$ and a lepton:
\begin{align}
N_{\tilde{\chi}^0_1}^{\mathrm{prod}}=\sum_M N_M\cdot \Gamma(M\rightarrow \tilde{\chi}^0_1 +l_i/\nu_i)\cdot \tau_M,
\end{align}
where $i=1, 2, 3$ is the lepton generation.

In principle one can consider the lightest neutralinos produced from either a pseudoscalar or a vector meson decay. However, the 
lifetime of the vector mesons is usually several orders of magnitude (up to 9 orders of magnitude for $D$-mesons for instance) 
lower than that of their pseudoscalar counterparts. In order to produce sufficiently many neutralinos from such vector meson decays 
mediated by an RPV coupling, the coupling has to be much larger than the current upper experimental bounds. Thus, we only 
consider pseudoscalar mesons for the neutralino production. Considering both $D$- and $B$-mesons rare decays, we are able to 
explore $\tilde{\chi}_1^0$ masses up to several GeV and down to a few hundred MeV. We follow the procedure given in 
Ref.~\cite{Dercks:2018eua} to extract the total number of the charm and bottom mesons that are relevant to this study, respectively. 
This is based on the experimental results of charm meson and $b-$quark production cross sections published by the LHCb collaboration 
\cite{Aaij:2015bpa, Aaij:2016avz}, and the kinematic extrapolation to the complete solid-angle coverage with the numerical tool \texttt{FONLL} \cite{Cacciari:1998it,Cacciari:2001td,Cacciari:2012ny,Cacciari:2015fta}, and $B$-meson fragmentation 
by using \texttt{Pythia 8} \cite{Sjostrand:2007gs,Sjostrand:2006za}. We summarize the results in 
Table~\ref{tab:Meson_Numbers}.\footnote{The numbers are slightly different from those given in  Ref.~\cite{Dercks:2018eua}. We have corrected
some minor errors in that paper.}

\begin{table}
	\begin{center}
		\begin{tabular}{c|c|c}
			\hline
			Meson $M$ &$D^\pm_s$ &   $B^0$/$\bar{B}^0$ \\
			\hline
			$N_M$  &  			  $6.62\times10^{15}$ &  $1.46\times10^{15}$ \\
			\hline 
		\end{tabular}
		\caption{The total number of $D$- and $B$-mesons expected at \texttt{ANUBIS} for an integrated luminosity of 3 ab$^{-1}$ 
		over the full solid angle $4\pi$. For \texttt{MAPP1} and \texttt{MAPP2} we scale the results to the respective 
		integrated luminosities, 30 fb$^{-1}$ and 300~fb$^{-1}$.}
		\label{tab:Meson_Numbers}
	\end{center}
\end{table}

The lightest neutralinos may undergo two-body decays into either charged or neutral final states. While both types contribute to the 
total decay width of $\tilde{\chi}_1^0$, only the charged final states can be easily used for the displaced-vertex reconstruction. We 
therefore consider only these as visible. The number of observed lightest neutralino decays can be expressed as
\begin{align}
N_{\tilde{\chi}_1^0}^{\mathrm{obs}}=N_{\tilde{\chi}_1^0}^{\mathrm{prod}}\cdot \Braket{P\left[\tilde{\chi}_1^0\text{ in d.r.}\right]}\cdot 
\mathrm{BR}\left(\tilde{\chi}_1^0\rightarrow \mathrm{char.}\right),
\end{align}
where  $\Braket{P\left[\tilde{\chi}_1^0\text{ in d.r.}\right]}$ denotes the average probability of the $\tilde{\chi}_1^0$ to decay inside 
the detectable region (d.r.) of a given detector and ``char.'' labels charged final states. We perform a Monte-Carlo (MC) simulation 
with \texttt{Pythia~8} in order to determine $\Braket{P\left[\tilde{\chi}_1^0\text{ in d.r.}\right]}$. We simulate $N_{\tilde{\chi}_1^0}^{\text{MC}}$ 
MC-events and calculate $\Braket{P\left[\tilde{\chi}_1^0\text{ in d.r.}\right]}$ with the following formula:
\begin{align}
\Braket{P\left[\tilde{\chi}_1^0\text{ in d.r.}\right]}=\frac{1}{N_{\tilde{\chi}_1^0}^{MC}}\sum_{i=1}^{N_{\tilde{\chi}_1^0}^{MC}}P\left[\left(\tilde{\chi}
_1^0\right)_i\text{ in d.r.}\right],
\end{align}
where $P\left[\left(\tilde{\chi}_1^0\right)_i\text{ in d.r.}\right]$ is the probability of an individual simulated $\tilde{\chi}_1^0$ to decay in the d.r.
The calculation of the individual decay probability takes into account the geometries of the respective detector and the kinematics 
of an individual $\tilde{\chi}_1^0$, and is explained in more detail below.

To perform the MC simulation, we use two modules, \texttt{HardQCD:hardccbar} and \texttt{HardQCD:hardbbbar} implemented in \texttt{Pythia 8}, 
in order to simulate $D$- and $B$-meson production in $pp-$collisions with the center-of-mass energy $\sqrt{s}=14$ TeV. For each parameter point, 
we simulate $2\times 10^6$ collisions for the bottom and $2\times 10^7$ collisions for the charm scenarios. We force the meson relevant for 
each benchmark scenario to exclusively decay into the lightest neutralino plus the accompanying lepton, in order to achieve the maximal number of 
statistics for estimating the 
average decay probability. We then compute the number of expected neutralino decays by including the total number of the mother meson produced 
and its decay branching ratio into $\tilde{\chi}_1^0$.

The calculation of the individual decay probabilities in the detector requires the kinematic information of each simulated neutralino. With the 
mass $m_{\tilde{\chi}^0_1}$ and the 4-momentum information ($E_i,$ $\theta_i,$ $\phi_i$) of the $i$-th simulated neutralino provided by 
\texttt{Pythia 8}, we can calculate the relativistic quantities with the following expressions:
\begin{align}
\gamma_i=&E_i/m_{\tilde{\chi}_1^0},\\
\beta_i=&\sqrt{\gamma_i^{2}-1}/\gamma_i,\\
\lambda_i=&\beta_i\gamma_i/\Gamma_{\text{tot}}(\tilde{\chi}_1^0),\\
\beta_i^z=&p_i^z/E_i, \\
\lambda_i^z=&\beta_i^z\gamma_i/\Gamma_{\text{tot}}(\tilde{\chi}_1^0),
\end{align}
where $\gamma_i$ is the Lorentz boost factor of the neutralino, $\beta_i$ ($\beta_i^z$) the relativistic speed (the velocity in the 
collider-beam direction), $\Gamma_{\mathrm{tot}}(\tilde{\chi}^0_1)$ the total decay width of the neutralino, $\lambda_i$ ($\lambda_i^z$) 
the boosted decay length in the traveling direction (in the beam direction).

\subsection{The Individual Decay Probability}

The traveling direction of an LLP is defined by the polar and azimuthal angles. $P\left[\left(\tilde{\chi}_1^0\right)_i\text{ in d.r.}\right]$ 
can then be estimated by
\begin{align}
 P\left[\left(\tilde{\chi}_1^0\right)_i\text{ in d.r.}\right]=
 e^{-\frac{L_{T,i}}{\lambda_i}}\left(1-e^{-\frac{L_{I,i}}{\lambda_i}}\right), \label{eqn:decayprobabilityindividualparticle}
\end{align}
in the case that the lightest neutralino travels inside the solid angle protruded by the detector. Otherwise the decay probability in the detector is 0. $L_{T,i}$ is the distance from the IP to the closest point of the detector, 
while $L_{I,i}$ is the distance the $i$-th simulated neutralino would travel inside the detector given its traveling direction, if it does not decay 
before it leaves the detector. Both $L_{T,i}=L_{T,i}(\theta_i,\phi_i)$ and $L_{I,i}=L_{I,i}
(\theta_i,\phi_i)$ are functions of the angles $\theta_i,\phi_i$ as well as the geometry of the detector at hand.

\subsubsection{\texttt{ANUBIS}}

In order to estimate the individual decay probability of an LLP inside the \texttt{ANUBIS} detector, we follow the same procedure as in 
Ref.~\cite{Hirsch:2020klk}. \texttt{ANUBIS} has 4 equally spaced tracking stations, between which we divide the detectable region into 
3 segments of height $l_v^\text{seg}=18.67$ m. For each of these regions we calculate separately $P_j\left[\left(\tilde{\chi}_1^0\right)_i
\text{ in d.r.}\right]$ with $j=1, 2, 3$ and then sum over all three probabilities. If one of the two following conditions is met
\begin{align}
	\tan{\theta_i} &\leq \frac{d_v+(j-1)\cdot l_v^\text{seg}}{d_h+l_h}  ,\\
	\tan{\theta_i}   &\geq  \frac{d_v+j\cdot l_v^\text{seg}}{d_h} ,
\end{align}
$P_j$ is 0. In the first case the neutralino flies below segment $j$, thus missing it. In the second case it flies above segment $j$. Otherwise 
we have
\begin{align}
	&P\left[\left(\tilde{\chi}_1^0\right)_i\text{ in d.r.}\right]=\sum_{j=1}^{3}\,\frac{\delta\phi^j}{2\pi}\,\cdot  e^{-\frac{L_{T,i}^j}{\lambda_i^z}} \cdot \left(1-e^{-\frac{L^{j}_{I,i}}{\lambda_i^z}}\right)\,,
	\label{eqn:decayprobabilityANUBIS}
\end{align}
where
\begin{align}	
	\delta\phi^j&=2\arctan{\frac{l_h/2}{d_v+(2j-1)/2\cdot l_v^{\text{seg}}}},\label{eqn:deltaphi}\\[2mm]
	L_{T,i}^j&=\text{min}\bigg[\text{max}\bigg(d_h,\frac{d_v+(j-1)\cdot l_v^\text{seg}}{\tan{\theta_i}}\bigg),d_h+l_h\bigg]\,,\\
	L_{I,i}^{j}&=\text{min}\bigg[\text{max}\bigg(d_h,\frac{d_v+j\cdot l_v^{\text{seg}}}{\tan{\theta_i}}\bigg),d_h+l_h\bigg]-L_i^j\,.
\end{align}
We do not determine the probability exactly for the azimuthal coverage. We assume the events are isotropic in $\phi$ and consider for each 
detector segment a cone around the $y$-axis to half the segment height. For the first segment we then have that $(\phi_i-\pi/2)\in[-\delta\phi,
+\delta\phi]$, where $\tan\delta\phi/2=(l_h/2)/(d_v+l_v^{\text{seg}}/2)$. Correspondingly for the other segments as in Eq.~(\ref{eqn:deltaphi}).
The azimuthal coverage is then accounted for with the prefactor $\delta\phi^j/2\pi$ in Eq.~(\ref{eqn:decayprobabilityANUBIS}).
Using this approximation for the location of the \texttt{ANUBIS} detector, and the probability for it to be hit by a particle flying from the IP,
we find a geometric coverage of the total solid angle of about 1.34\%.\footnote{The solid angle coverage is determined using Monte-Carlo 
integration with $10^6$ events. We display the mean solid angle, $\bar{\Omega}$, of 100 such integrations. The relative standard deviation 
($\sigma/\bar{\Omega}$) is $<0.3\%$ for all mentioned solid angles.}

\subsubsection{\texttt{MAPP}}

Because of the less regular orientation of the \texttt{MAPP} detectors, it is not straightforward to compute the individual decay probability, as for 
\texttt{ANUBIS}. Instead, we simulate the \texttt{MAPP} detectors in an exact way in three-dimensional space. For this we construct a 
virtual model of the detectors based on their corner points. This defines the surfaces of the detectors and thus the entire volume. Using the 
information of the polar and azimuthal angles of each simulated neutralino, our program determines whether the neutralino is traveling in a 
direction inside the detector window, and if so computes the $L_{T,i}$ and $L_{I,i}$ as given in Eq.~\eqref{eqn:decayprobabilityindividualparticle}, 
with which the individual decay probability in the \texttt{MAPP} detectors can be exactly determined. Using this more precise method, we find 
that \texttt{MAPP1} and \texttt{MAPP2} geometrically cover about 0.17\% and 0.68\% of the total solid angle, respectively.

We note that in principle this method can also be used for a cylindrical detector such as \texttt{ANUBIS}. However, given the rather small 
azimuthal-angle coverage of the \texttt{ANUBIS} detector, the amount of simulation increases drastically in order to reduce the numerical uncertainty 
to a sufficiently small level. Therefore, we use Eq.~\eqref{eqn:decayprobabilityANUBIS}, which does not require a too large number of simulated events
and at the same time is a sufficiently good approximation. With this exact geometry of the \texttt{ANUBIS} detector, the solid-angle 
coverage is estimated to be 1.79\%, which is slightly larger than that obtained by the approximate geometry (see Eq.~\eqref{eqn:decayprobabilityANUBIS}).
Thus our previous estimate is conservative. The effect on the sensitivity in each RPV coupling enters via the square root, thus the effect is
about 15\% in each coupling.

\section{Numerical Results for the benchmark scenarios}
\label{sect:results}

Here we present our numerical results. In Refs.~\cite{deVries:2015mfw,Dercks:2018eua} a variety of benchmark 
scenarios involving different $LQ\bar{D}$ couplings were investigated. These scenarios consider the production 
of light neutralinos via the rare decay of charged or neutral charm and bottom mesons. The neutralinos subsequently 
decay into lighter mesons. In each scenario, two $LQ\bar{D}$ couplings are switched on, one responsible for the 
production  and one for the decay of the neutralinos. In this work, we focus on only two specific benchmark scenarios, 
one for charmed and one for bottom mesons.

The relevant matrix elements for the production and decay of the neutralinos are given in Ref.~\cite{deVries:2015mfw}.
The effective production and decay operators are proportional to the RPV-couplings scaled to the squared sfermion mass $\lambda'/m^2_{\tilde{f}}$.
This is in contrast to the existing low-energy bounds listed in Eqs.~\eqref{eqn:singleRPVbounds1}-\eqref{eqn:RPVproductbounds2}, which scale with the sfermion masses as $\lambda'^2/m^2_{\tilde{f}}$ or $\lambda'/m_{\tilde{f}}$, for reasons explained in Sec~\ref{sect:theory}.
For simplicity, we assume degenerate sfermion masses, so that we are left with three free parameters for each benchmark scenario: the scaled production and decay couplings $\lam'_{P,ijk}/m_{\tilde{f}}^2$ and $\lam'_{D,i'j'k'}/m_{\tilde{f}}^2$, and the neutralino mass $m_{\tilde{\chi}^0_1}$.
With three independent parameters, we choose to present the sensitivities in two types of parameter planes.
First, we set the two $\lam'$-couplings to be equal and lay out the dependence on the neutralino mass: $\lam'_{P,ijk}/m_{\tilde{f}}^2=\lam^\prime_{D,i'j'k'}/m_{\tilde{f}}^2$ vs. $m_{\tilde{\chi^0_1}}$. The other parameter plane chosen is $\lambda^\prime_{P,ijk}/m_{\tilde{f}}^2$ vs. $\lam^\prime_{D,i'j'k'}/m_{\tilde{f}}^2$ for three fixed neutralino masses, where we vary the two $LQ\bar{D}-$couplings independently.

We further display the sensitivities in the plane Br vs. $c\tau$, where Br denotes the decay branching ratio of the mother 
meson times that of the neutralino into charged products, and $c\tau$ is the proper decay length of the neutralino. If the 
decay topologies are similar, these results should not be different qualitatively in the context of other theoretical models. 
As mentioned, we consider only the charged final states can be detected and used for the DV reconstruction.

\subsection{Benchmark scenario 1 - charmed Meson $D_s$}

\begin{table}[t]
	\centering
	\begin{tabular}{l|l}
		\hline
		\hline
		$\lambda^\prime_P$ for production & $\lambda^\prime_{122}$\\ 
		$\lambda^\prime_D$ for decay  & $\lambda^\prime_{112}$\\
		produced meson(s) & $D_s$\\
		visible final state(s) & $K^\pm +e^\mp$, $K^{*\pm}+e^\mp$\\
		invisible final state(s) via $\lambda^\prime_P\hspace{5mm}$ & $\left(\eta,\eta^\prime,\phi\right)+\left(\nu_e,\overline{\nu}_e\right)$\\
		invisible final state(s) via $\lambda^\prime_D\hspace{5mm}$ & $\left(K^0_L,K_S^0,K^*\right)+\left(\nu_e,\overline{\nu}_e\right)$\\
		\hline
		\hline
	\end{tabular}
	\caption{Features of the charmed benchmark scenario.}\label{tab:benchmark_scenario_charm}
\end{table}

For the first scenario we consider $\lambda'_{122}$ and $\lambda'_{112}$ to be non-zero, mediating the production 
and decay of the lightest neutralino, respectively. We start with a $D_s-$meson decaying promptly via the $L_1 Q_2 
\bar{D}_2$ operator:
\begin{equation}
D_s \rightarrow \tilde{\chi}^0_1 +e^\pm.
\end{equation}
Afterwards the neutralino travels a macroscopic distance before decaying with a displaced vertex via $\lambda'_{112}$ into either charged or neutral states:
\begin{equation}
\tilde{\chi}^0_1\rightarrow \left\{   \begin{array}{l}
K^{(*)\pm} + e^\mp\,,\\[1.5mm]
K^0_{L/S}/K^{*0}+\nu_e\,.
\end{array}     \right. \label{eq:BM1-vis-decay}
\end{equation}
In addition the production coupling induces invisible neutralino decays:
\begin{equation}
\tilde{\chi}^0_1\rightarrow \left\{   \begin{array}{l}
\eta/ \eta^\prime +\nu_e\,,\\
\phi +\nu_e\,,
\end{array}     \right. \label{eq:BM1-inv-decay}
\end{equation}
which must be taken into account when computing the total decay width.
The relevant features of benchmark scenario 1 are summarized in Table~\ref{tab:benchmark_scenario_charm}.

\begin{figure*}[t]
	\centering
	\includegraphics[height=.22\paperheight , width=.45\linewidth]{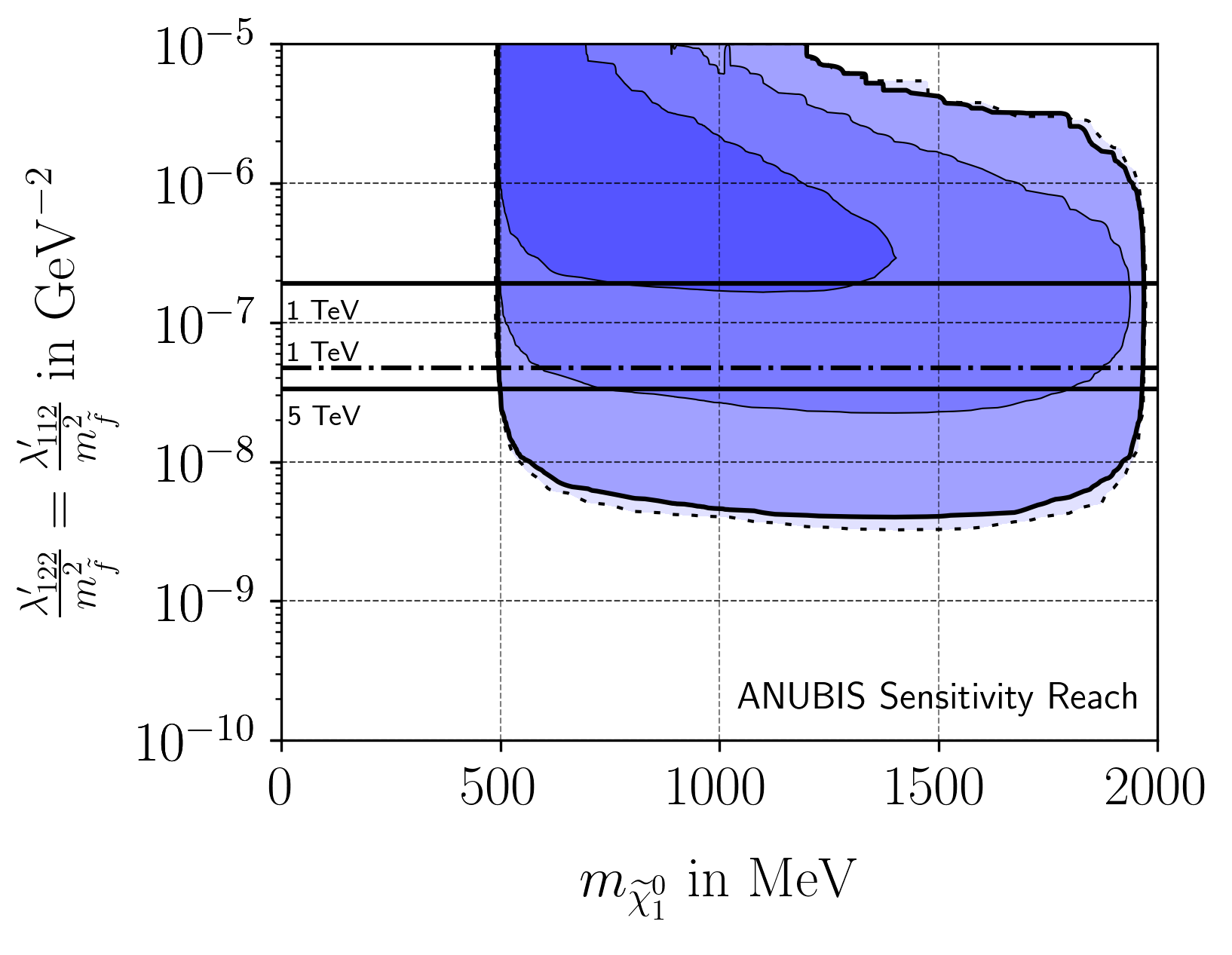}
	\includegraphics[height=.22\paperheight , width=.45\linewidth]{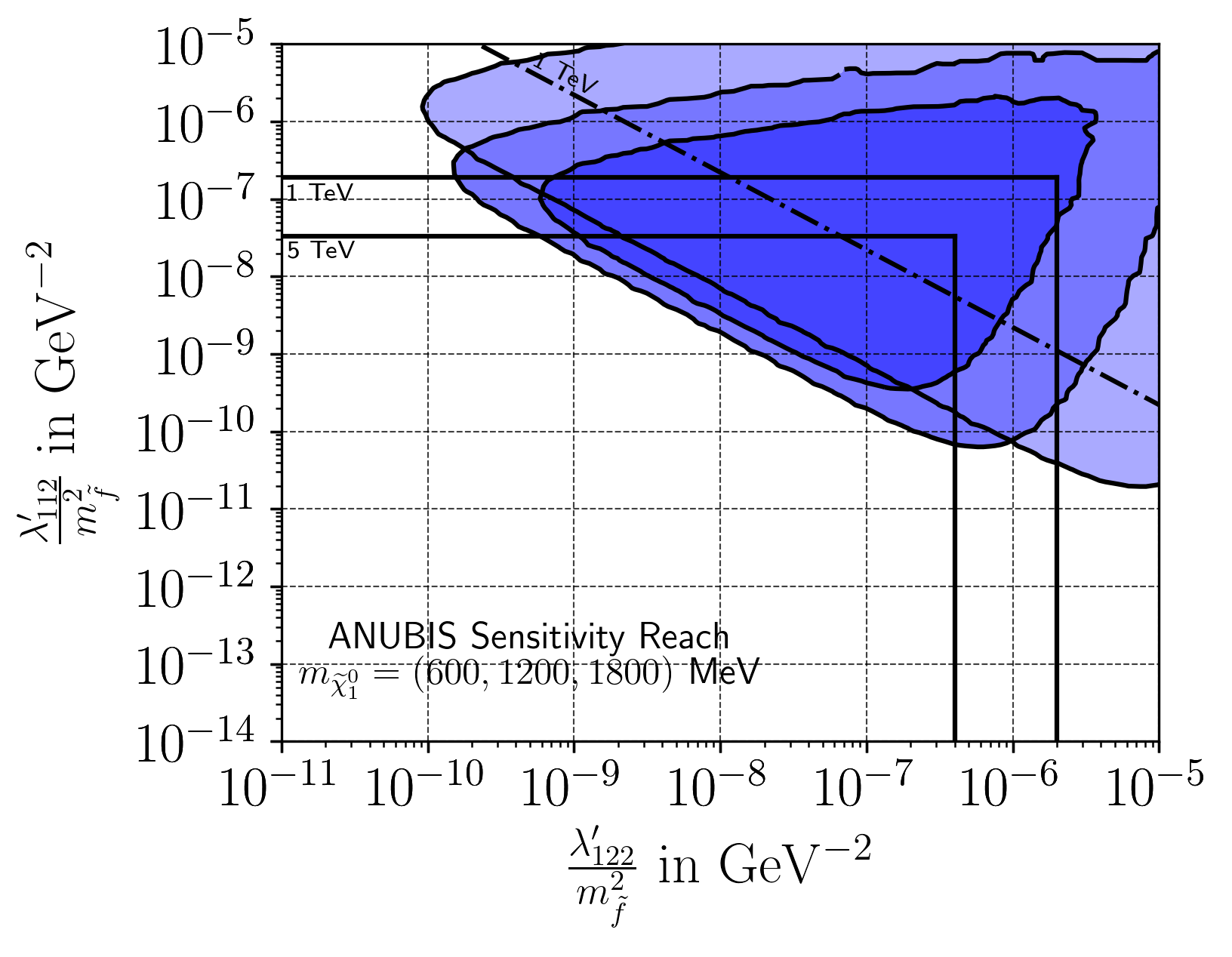}\\
	\includegraphics[height=.22\paperheight , width=.45\linewidth]{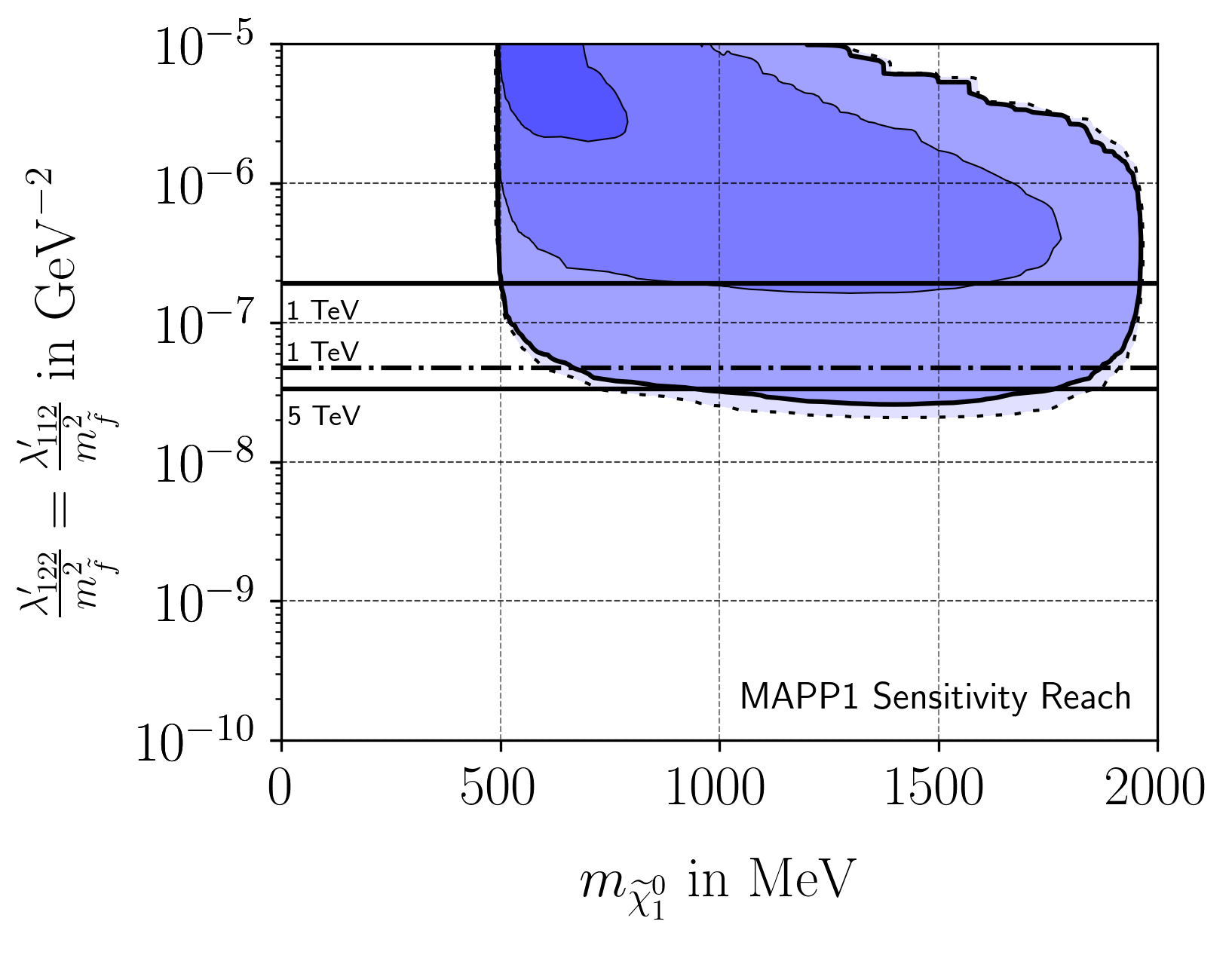}
	\includegraphics[height=.22\paperheight , width=.45\linewidth]{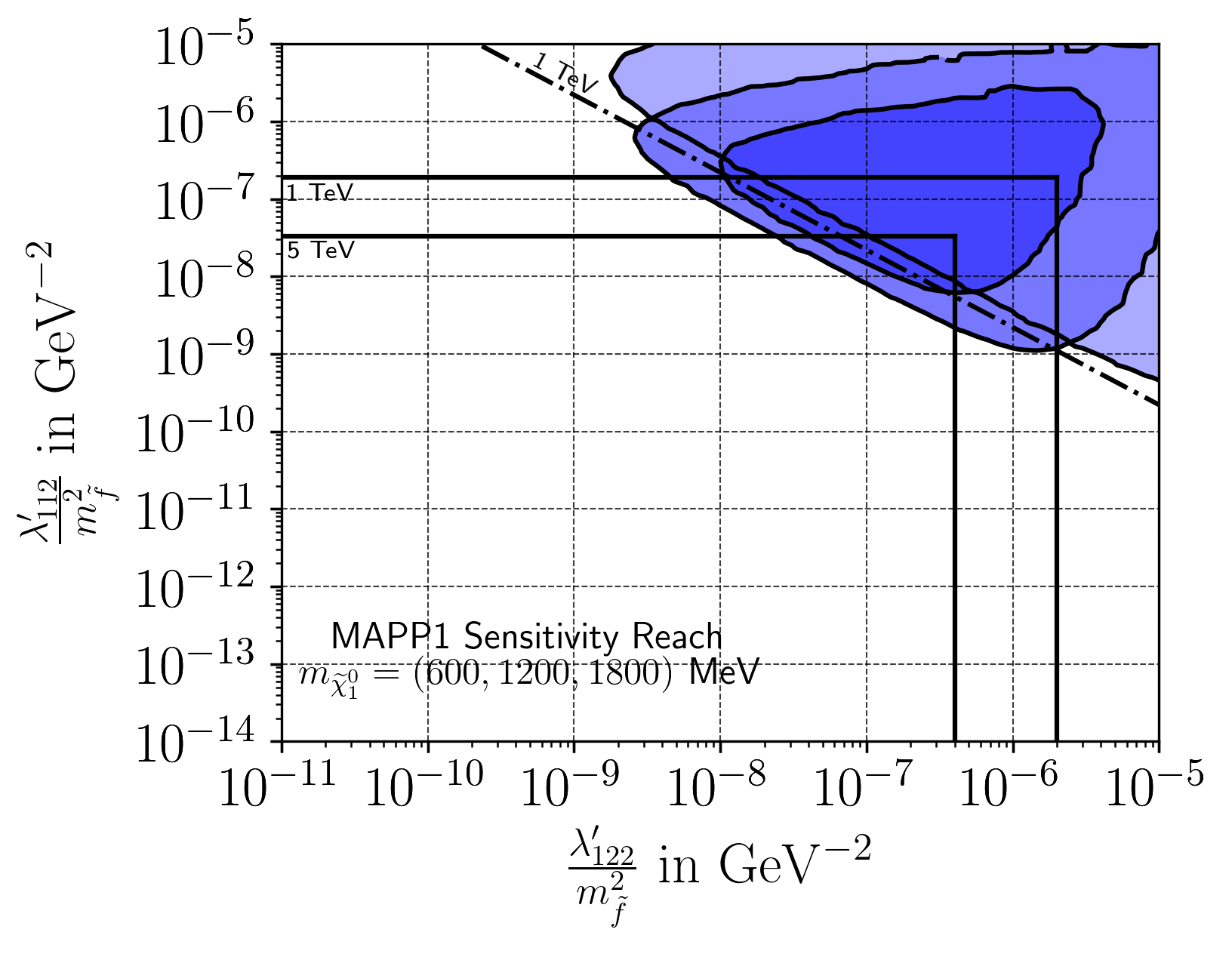}\\
	\includegraphics[height=.22\paperheight , width=.45\linewidth]{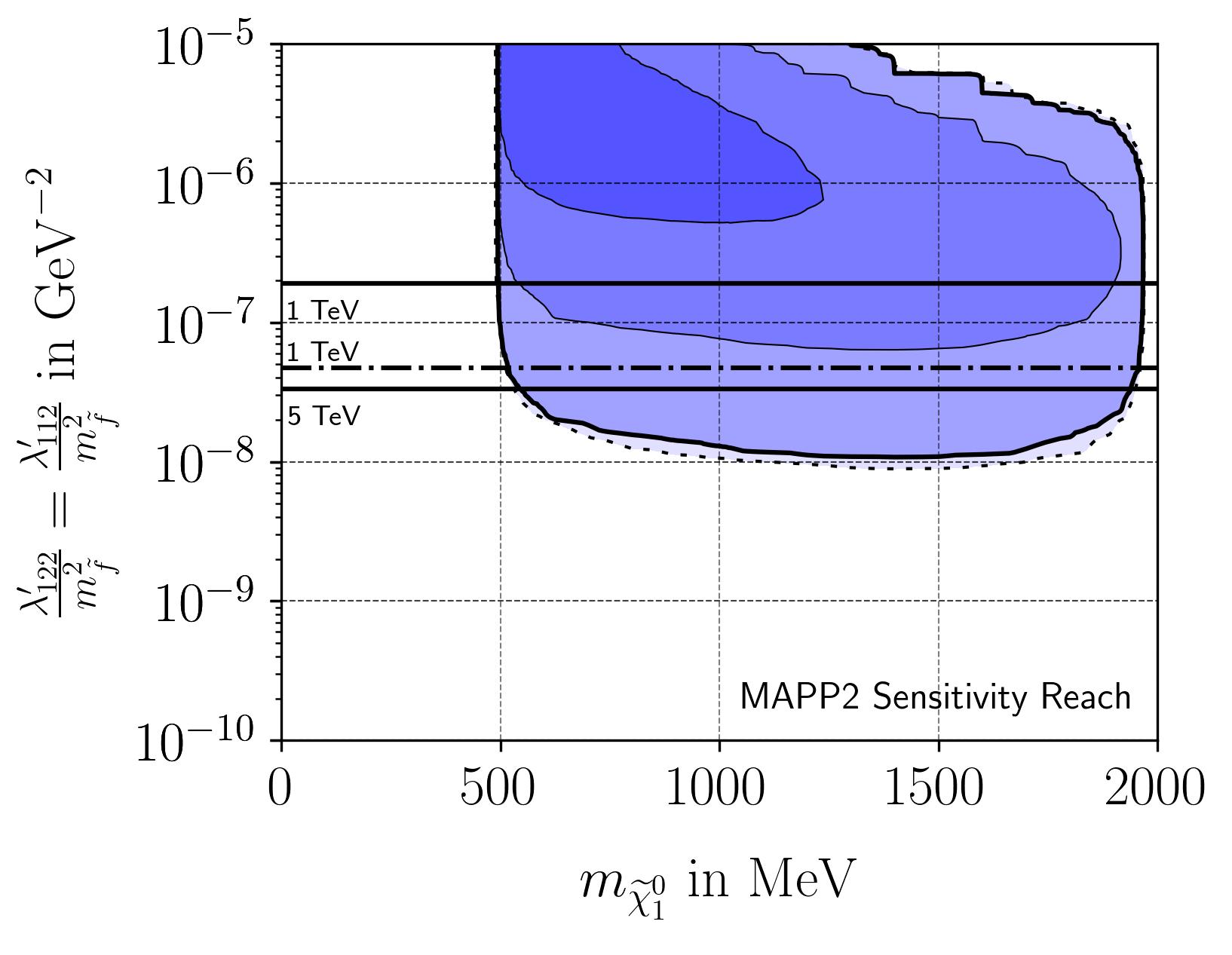}
	\includegraphics[height=.22\paperheight , width=.45\linewidth]{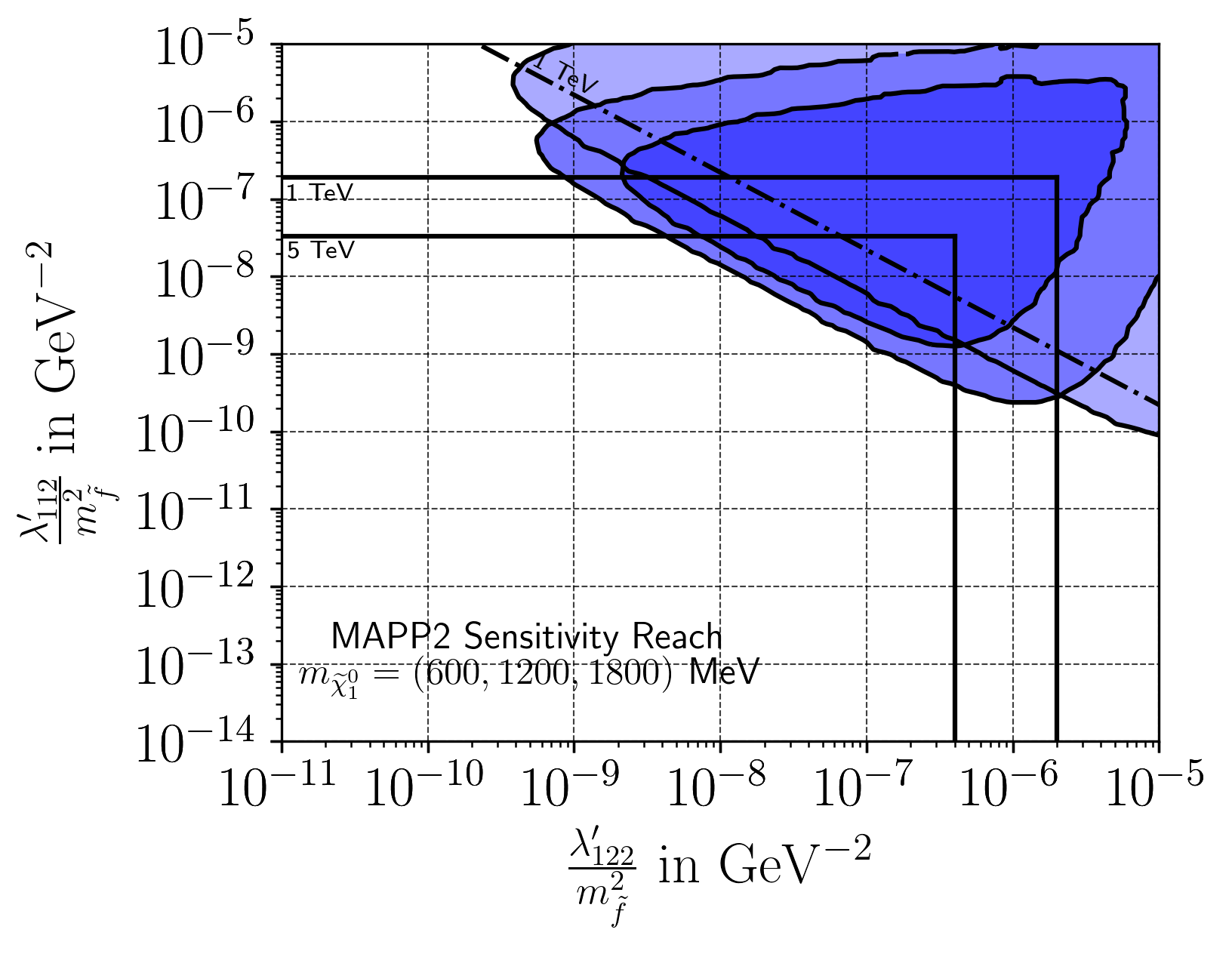}
	\caption{Estimated sensitivity reach for \texttt{ANUBIS}, \texttt{MAPP1}, and \texttt{MAPP2} for benchmark 
	scenario 1, with charmed mesons.  For the left column we demand the two $LQ\bar{D}$ couplings to be equal 
	and detail the reach as a function of the neutralino mass. The isocurves represent 3 events (light blue), $3\times 
	10^3$ events (medium blue),  and $3\times 10^6$ (dark blue). The dashed isocurve	is an extension of the 3-event 
	isocurves, if the neutral `invisible' decays can be observed in the detector. We implement the stronger current 
	coupling bound (here for $\lam^\prime_{112}$) for two different sfermion masses, $m_{\tilde{f}}=1\,$TeV and 
	$5\,$TeV (solid), and the product bound for $m_{\tilde{f}}=1\,$TeV (dot-dashed). Depicted in the right column are 
	3-event isocurves on the $\lam'_D$ vs. $\lam'_P$ parameter region for 3 specific neutralino masses, namely 
	$600\,$MeV (light blue), $1200\,$MeV (medium blue), and $1800\,$MeV (dark blue). Current upper limits on the 
	individual couplings for two sfermions masses $m_{\tilde{f}}=1\,$TeV and $5\,$TeV (solid), as well as the product 
	limit for the sfermion mass $m_{\tilde{f}}=1$ TeV (dot-dashed) are presented.	}\label{fig:122-112-model-dep}
\end{figure*}

In Fig.~\ref{fig:122-112-model-dep} we present the model-dependent results for this benchmark scenario.  The left 
column contains plots in the plane $\lam^\prime/m_{\tilde{f}}^2$ vs. $m_{\tilde\chi^1_0}$ for \texttt{ANUBIS}, 
\texttt{MAPP1}, and \texttt{MAPP2}  experiments, respectively, where we impose $\lam'_{122}=\lambda'_{112}$. In 
these plots we show contours of three different numbers of signal events with the light, medium, and dark blue areas 
corresponding to parameter regions with $>\!3$, \mbox{$>\!3\times 10^3$}, and $>\!3\times 10^6$ signal events, 
respectively. Furthermore, the 3-event isocurve is extended by the dashed line which gives the sensitivity reach if the 
neutral final states, the lower set of decays in Eq.~(\ref{eq:BM1-vis-decay}) and the decays in Eq.~(\ref{eq:BM1-inv-decay}), can be detected. Current bounds on the 
RPV-couplings as given in Eq.~\eqref{eqn:singleRPVbounds1} (Eq.~\eqref{eqn:RPVproductbounds1}) are shown with 
solid (dot-dashed) horizontal lines. For the single coupling bounds on $\lambda'_P$ and $\lambda'_D$, we show only 
the stronger one for sfermion masses of 1\,TeV and 5\,TeV, while for the bound on the product of the two RPV couplings ($\sqrt{\lambda'_{122}\lambda'_{112}}/m^2_{\tilde{f}}$) we consider only one sfermion mass at 1 TeV.

The sensitive neutralino mass range for these experiments are all similar and constrained mainly by the kinematics of 
the scenario:
\begin{equation}
\left(M_{K^\pm}+m_e\right)< m_{\tilde{\chi}^0_1}<\left(M_{D_s}-m_e\right).
\end{equation}
Beyond this, the sensitivities are dependent only to a small extent on the neutralino mass, depicted here by the slope  
of the lower edge of the various sensitivity regions in the plots in the left column. The sensitivity regions are bounded 
from above, as for large couplings the neutralinos would decay too fast to reach the detector. They are bounded from 
below since for small couplings there is both insufficient production of the neutralinos and a too large decay length. The 
slope of the upper edge of the sensitivity regions can be understood as follows. Increasing the two RPV couplings 
and decreasing the neutralino mass simultaneously, the observed number of signal events increase in general. This is 
because a large $\lam'_P$ leads to enhancement in the neutralino production, while the increase in $\lam'_D$ and 
decrease in $m_{\tilde{\chi}_1^0}$ retains the decay width and the average decay probabilities in the detector.

In benchmark scenario 1, all three experiments may probe parameter regions beyond the current RPV-coupling bounds, 
to different extents. While the sensitivity of  \texttt{MAPP1} beyond the current limits would be less than a factor 2 in 
$\lam'/m^2_{\tilde{f}}$, its upgraded version, \texttt{MAPP2}, may extend the reach of \texttt{MAPP1} by a further factor of
$\sim 3$, by virtue of its larger volume and the increased  integrated luminosity. Among the three experiments 
studied in this work, \texttt{ANUBIS} shows the best sensitivity reach, exceeding the current limits by a factor $\sim 8$ in 
$\lam'/m^2_{\tilde{f}}$. This can be attributed to its even greater integrated luminosity and a larger solid-angle coverage.

The plots in the right column of Fig.~\ref{fig:122-112-model-dep} are shown in the plane $\lambda^\prime_D/m_{\tilde{f}}^2$ 
vs. $\lam^\prime_P/m_{\tilde{f}}^2$. 3-event isocurves in different colors are presented for three fixed neutralino masses: 
600 (light blue), 1200 (medium blue), and 1800 MeV (dark blue). These choices of the neutralino mass correspond 
approximately to the lower and higher ends, and the middle point of the mass range allowed by the kinematics, as discussed 
above. As in the plots on the left, we included present experimental bounds on both single couplings and  the couplings'
product for different sfermion masses.

The isocurves are bounded from all four sides. With a too large/small $\lambda'_D$ ($y$-axis), the lightest neutralinos 
decays too early/late, leading to a too small average decay probability in the detectors. When $\lambda'_P$ ($x$-axis) 
is too small, there is insufficient production of the neutralinos. With a too large $\lambda'_P$, the lightest neutralinos 
would also decay before they reach the detector. This is specific for this scenario, as $\lam'_P$ also 
induces invisible decays of the neutralino and hence enhances its total decay width.

We find that \texttt{ANUBIS}, \texttt{MAPP1}, and \texttt{MAPP2} can all be sensitive to new parameter regions beyond 
the current RPV limits for sfermion masses of the order of $1\,$TeV. As observed in the plots in the left column, compared 
to \texttt{MAPP1}, \texttt{MAPP2} shows better sensitivity reach while \texttt{ANUBIS} is expected to have the strongest 
performance. Assuming $m_{\tilde{f}}=5\,$TeV as a reference value, \texttt{MAPP1} improves the current bounds on 
$\lam'_{122}$ and $\lam'_{112}$ by approximately one order of magnitude, whereas \texttt{MAPP2}  and \texttt{ANUBIS} 
improve them by more than two orders of magnitude. In general among the three masses considered, $m_{\tilde{\chi}_1^0}
=1200\,$MeV probes the largest part of the parameter regions that are still allowed by the present limits.

\begin{figure}[t]
	\centering
	\includegraphics[width=.999\linewidth]{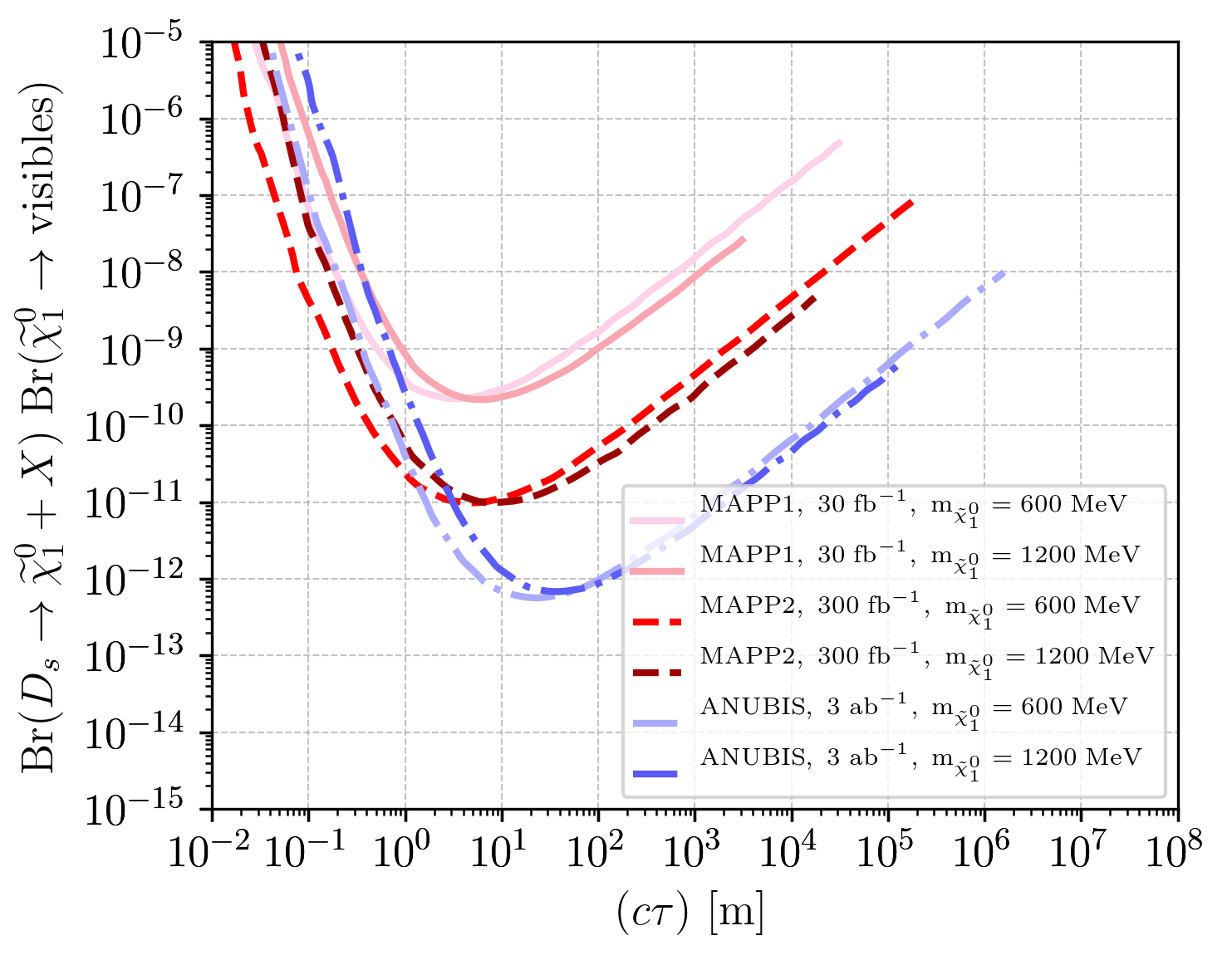}
	\caption{Sensitivity estimates for a neutral long-lived fermion in a model-independent description for the charmed benchmark scenario 1 using our 
	neutralino estimates. For each detector the estimates for two light neutralino masses are depicted. For benchmark scenario 1 these masses  are 
	$600$ and $1200\,$MeV. The lighter colors are used for the lighter mass. \texttt{MAPP1} is illustrated with a pink solid curve, \texttt{MAPP2} red 
	with a dashed line and \texttt{ANUBIS} blue with a dot-dashed line.}\label{fig:122-112-model-indep}
\end{figure}

In Fig.~\ref{fig:122-112-model-indep}, we show the model-independent results for benchmark scenario 1 in the plane Br vs. $c\tau$ for two neutralino masses.
Here we specify ``Br'' with the following expressions:
\begin{eqnarray}
\text{Br}&\equiv&\text{Br}\left(D_s^\pm  \rightarrow \tilde{\chi}_1^0 +  e^\pm \right) \cdot \text{Br}\left( \tilde{\chi}_1^0    \rightarrow K^{(*)\pm}+e^\mp\right)\!. \ \ \ \ 
\end{eqnarray}
The solid pink curves denote \texttt{MAPP1}, the dashed red curves \texttt{MAPP2}, and the dot-dashed blue curves 
\texttt{ANUBIS}. The lighter mass $600\,$MeV correlates to the lighter color, while the darker color is for a mass of 
$1200\,$MeV. The relative comparison between these experiments is similar to that shown in 
Fig.~\ref{fig:122-112-model-dep}. The minimum of each curve gives the lowest reach in Br, and its corresponding 
position in the proper lifetime. For this scenario, \texttt{ANUBIS} may reach $\mathrm{Br}\, \approx \,7\times 10^{-13}$, 
while the two \texttt{MAPP} programs are {less sensitive by more than one order of magnitude.}

\begin{table}
	\centering
	\begin{tabular}{lc||ccc}
		\hline\hline
		Scenario & $m_{\tilde{\chi}^0_1}$ [MeV] & $\Braket{\beta\gamma}_{\mathtt{ANUBIS}}\;$& $\Braket{\beta\gamma}_{\mathtt{MAPP1}}\;$ & $\Braket{\beta\gamma}_{\mathtt{MAPP2}}$\\
		\hline
		1 - $D_s$ & 600  & 2.78 & 24.45  & 17.22\\
		1 - $D_s$ & 1200 & 2.94 & 16.63 & 13.26  \\
		2 - $B^0/\overline{B}^0$ & 1000 & 5.56 & 37.86 & 26.77  \\
		2 - $B^0/\overline{B}^0$ & 3000  & 2.62 & 16.42 & 12.72 \\
		\hline\hline
	\end{tabular}
	\caption{Average boost factors of the neutralinos for the three detectors and both benchmark scenarios.
			 }\label{tab:average_neutralino_boost}
\end{table}

In order to understand the resulting lowest points in Br: $(c\tau)_{\text{min}}$, in Fig.~\ref{fig:122-112-model-indep}, 
we first consider the distance from the IP to the middle position of the detector, $\Braket{L}$. For the three detectors 
we have
\begin{equation}\label{eqn:avg_det_lengths}
\Braket{L}=\left\{  \begin{matrix}
\hspace{2mm}53.85\text{ m}&\hspace{3.9mm}\text{ for \texttt{ANUBIS},}\\
\hspace{2mm}55.06\text{ m}&\hspace{2mm}\text{ for \texttt{MAPP1},}\\
\hspace{2mm}46.71\text{ m}&\hspace{2mm}\text{ for \texttt{MAPP2}.}\\
\end{matrix}\right.
\end{equation}
We estimate the average boost $\Braket{\beta\gamma}$ of the produced neutralinos for each detector with our numerical simulation, 
with the results listed in Table~\ref{tab:average_neutralino_boost}. We note that for the two \texttt{MAPP} programs the relevant neutralinos have similar average boost factors while the neutralinos traveling inside the \texttt{ANUBIS} window have 
an average boost factor approximately one order of magnitude smaller. This is mainly due to the difference in the polar-angle coverage 
of these experiments. Since at the LHC the charm and bottom mesons that would decay to the light neutralinos are on average highly boosted in the forward direction, the neutralinos that travel at a larger polar angle tend to be less energetic. As \texttt{ANUBIS} 
covers polar angles larger than \texttt{MAPP1} and \texttt{MAPP2}, the neutralinos going into \texttt{ANUBIS} are expected to have a 
smaller boost factor. 

The  minima of the curves are found to correspond to
\begin{equation}
(c\tau)_{\text{min}}\approx \Braket{L}/\Braket{\beta\gamma}. \label{eqn:ctaumax}
\end{equation} 
This corresponds to the point of highest sensitivity to the product of branching ratios.
Taking $m_{\tilde\chi^0_1}=600\,$MeV as an example, we obtain
\begin{equation}\label{eqn:valley_positions_charmed}
(c\tau)_{\mathrm{min}}=\left\{  \begin{matrix}
\hspace{2mm} 19.37 \text{ m}&\hspace{3.5mm}\text{ for \texttt{ANUBIS},}\\
\hspace{2mm} 2.25 \text{ m}&\hspace{2mm}\text{ for \texttt{MAPP1},}\\
\hspace{2mm} 2.71 \text{ m}&\hspace{2mm}\text{ for \texttt{MAPP2},}
\end{matrix}\right.
\end{equation}
in approximate agreement with the minima of the lighter-colored curves in Fig.~\ref{fig:122-112-model-indep}.
The value of (Br)$^2$ at $(c\tau)_{\text{min}}$ depends on the angular coverage of the detector and the integrated luminosity.

\begin{figure*}[t]
	\centering
	\includegraphics[height=.22\paperheight , width=.45\linewidth]{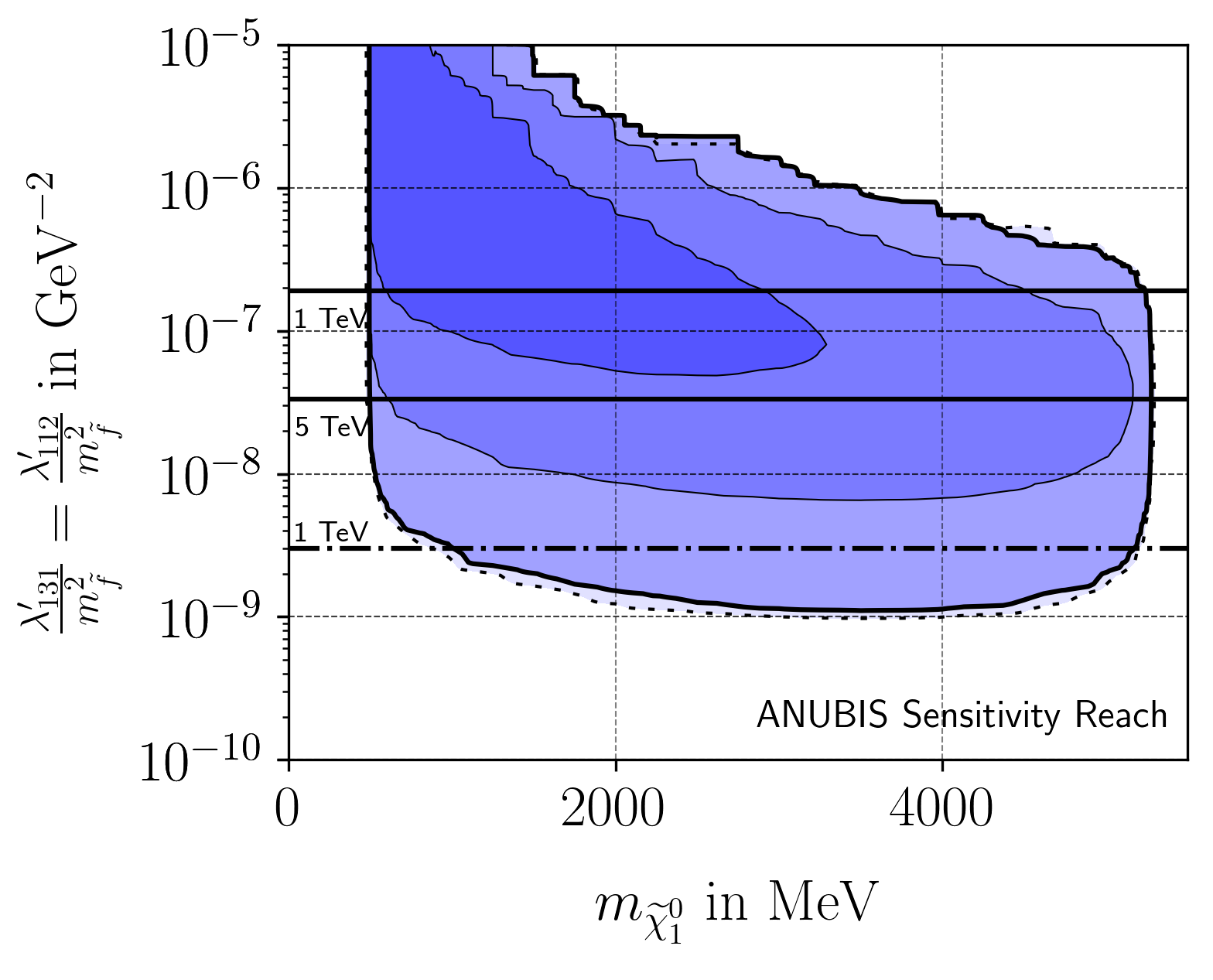}
	\includegraphics[height=.22\paperheight , width=.45\linewidth]{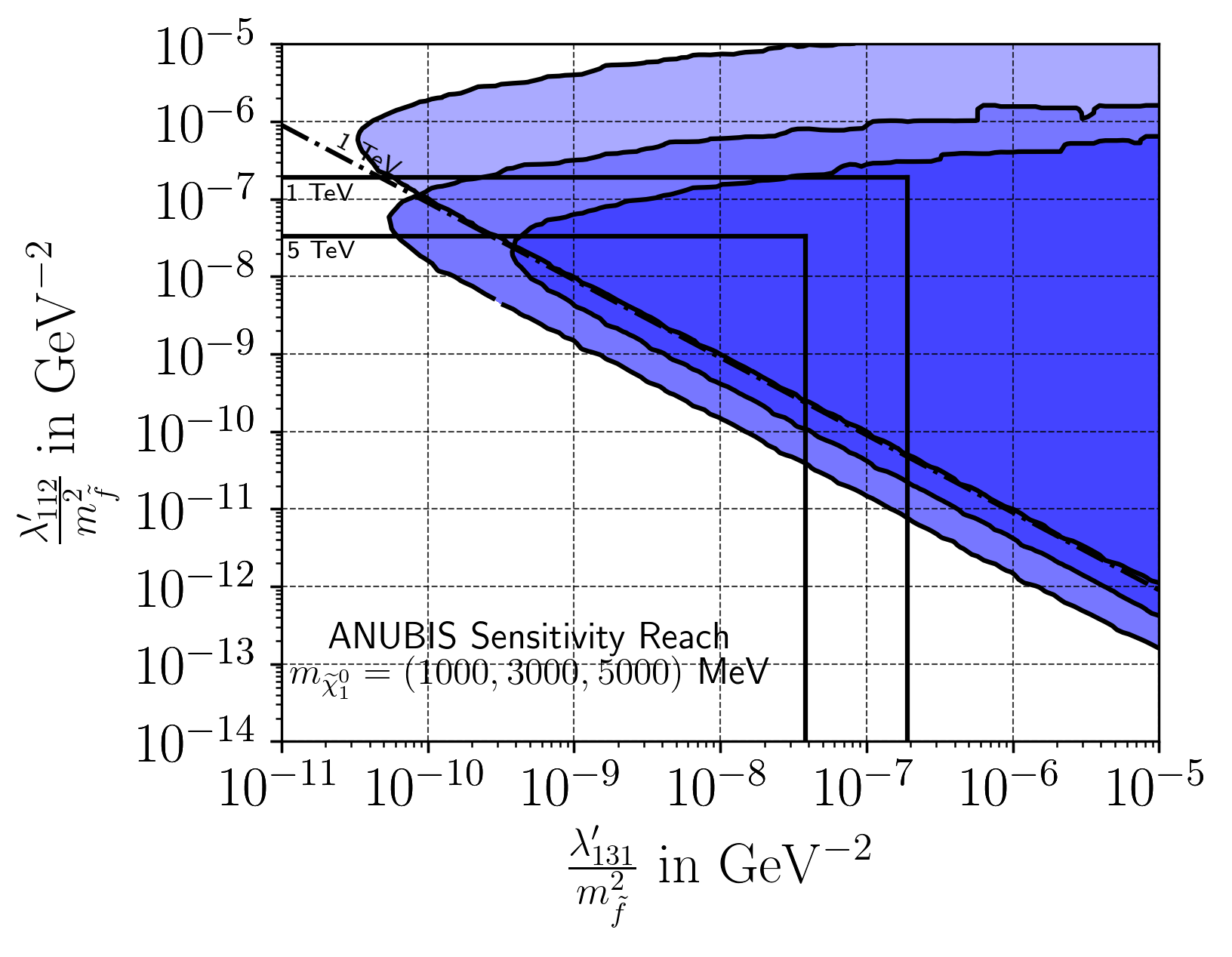}\\
	\includegraphics[height=.22\paperheight , width=.45\linewidth]{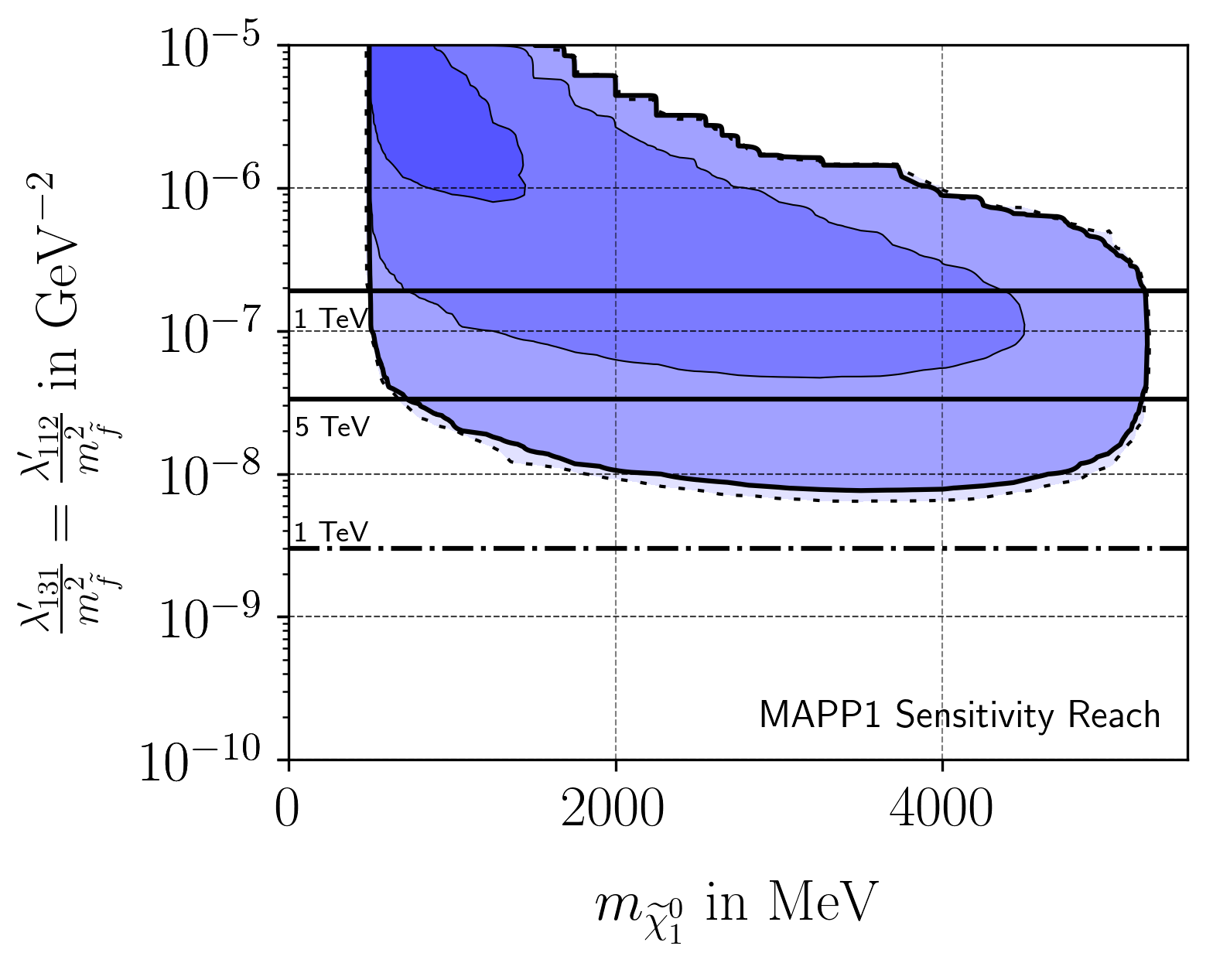}
	\includegraphics[height=.22\paperheight , width=.45\linewidth]{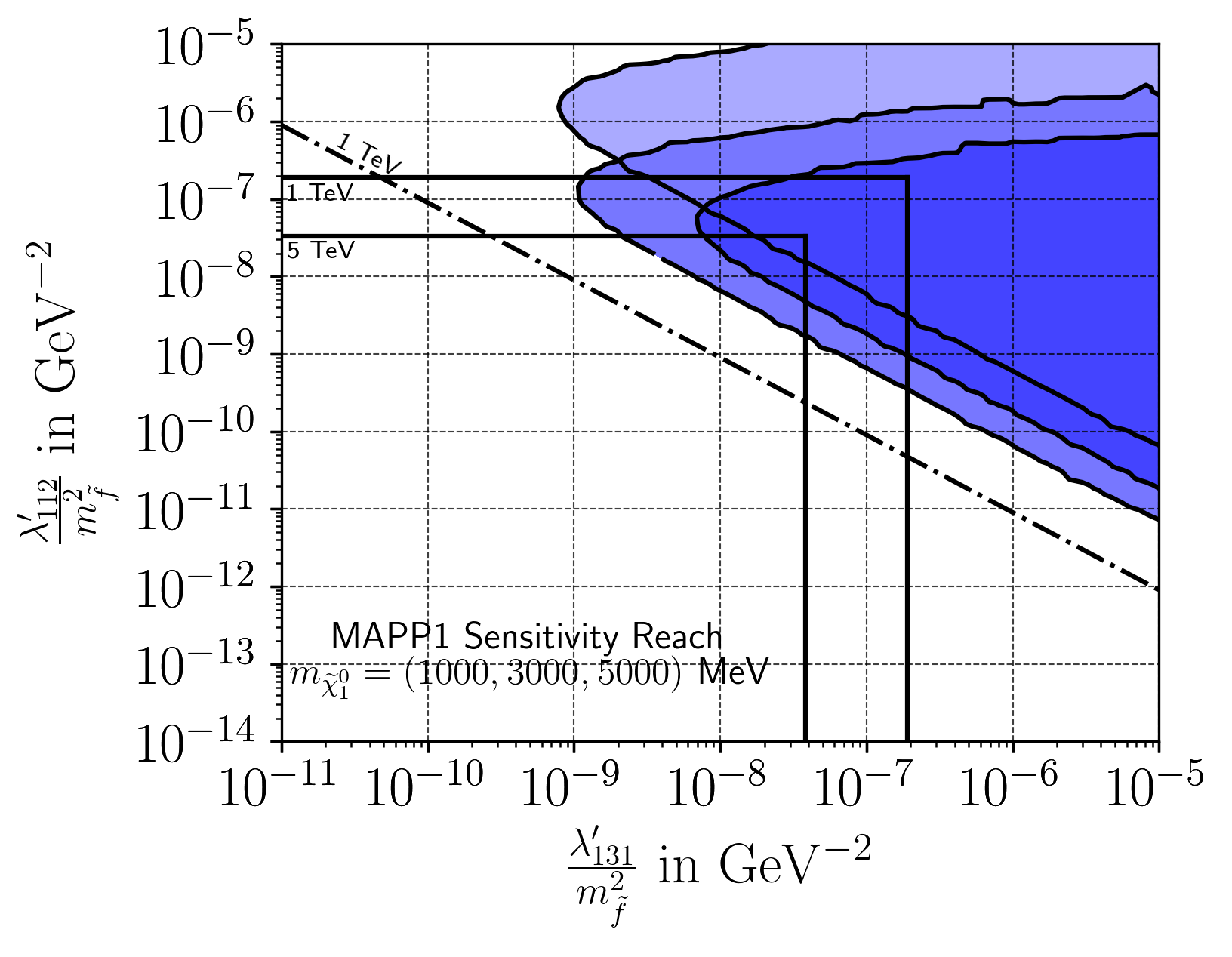}\\
	\includegraphics[height=.22\paperheight , width=.45\linewidth]{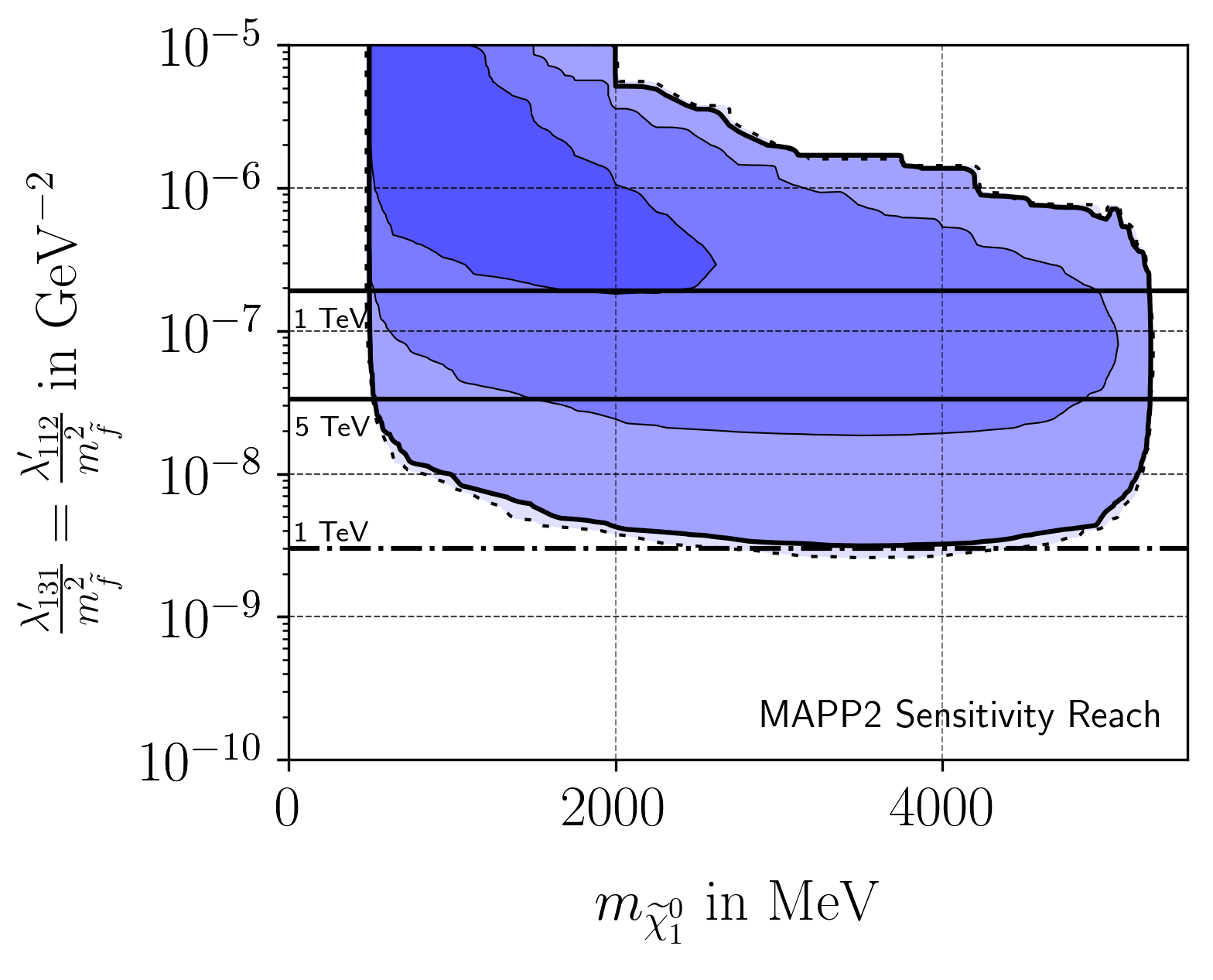}
	\includegraphics[height=.22\paperheight , width=.45\linewidth]{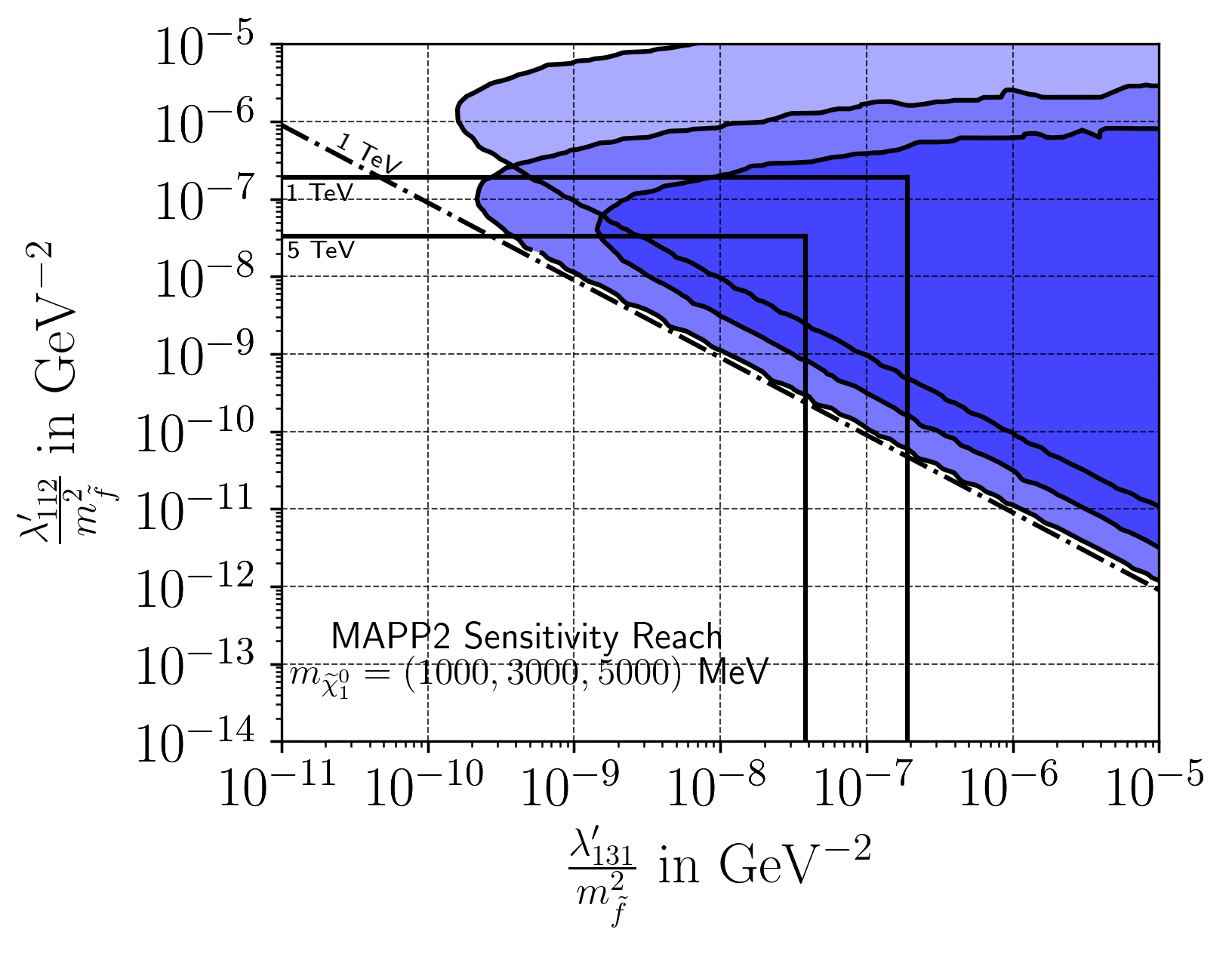}
	\caption{Estimated sensitivity reach for \texttt{ANUBIS}, \texttt{MAPP1}, and \texttt{MAPP2} in the $\lam^\prime_{131}/m_{\tilde{f}}^2$ vs. $\lam^\prime_{112}/m_{\tilde{f}}^2$ parameter plane for the bottomed benchmark scenario.
		The labeling is similar to that in Fig.~\ref{fig:122-112-model-dep}, whereas in the right column the neutralino masses considered now are $m_{\tilde{\chi}^0_1}=1000$ MeV, 3000 MeV, and 5000 MeV colored as light blue, medium blue, and dark blue, respectively. }\label{fig:131-112-model-dep}
\end{figure*}

\subsection{Benchmark scenario 2 - bottomed Meson $B^0,\overline{B}^0$}

\begin{table}[t]
	\centering
	\begin{tabular}{l|l}
		\hline\hline
		$\lambda^\prime_P$ for production & $\lambda^\prime_{131}$\\ 
		$\lambda^\prime_D$ for decay  & $\lambda^\prime_{112}$\\
		produced meson(s) & $B^0$, $\overline{B}^0$\\
		visible final state(s) & $K^\pm +e^\mp$, $K^{*\pm}+e^\mp$\\
		invisible final state(s) via $\lambda^\prime_P\hspace{5mm}$ & None\\
		invisible final state(s) via $\lambda^\prime_D\hspace{5mm}$ & $\left(K^0_L,K_S^0,K^*\right)+\left(\nu_e,\overline{\nu}_e\right)$\\
		\hline\hline
	\end{tabular}
	\caption{Features of the bottomed benchmark scenario.}\label{tab:benchmark_scenario_bottom}
\end{table}
Next, we consider neutral bottom mesons $B^0$ decaying to a neutralino plus a neutrino via the coupling $\lambda'_{131}$.
The decay of the lightest neutralino into a kaon proceeds via the same coupling $\lambda'_{112}$ as that considered in the previous benchmark scenario.
The characterizing features of this benchmark are summarized in Table~\ref{tab:benchmark_scenario_bottom}.
This extends the mass range for the neutralino considerably because of the larger mass of the B-meson:
\begin{equation}
\left(M_{K^\pm}+m_e\right)< m_{\tilde{\chi}^0_1}<\left(M_{B^0}-m_{\nu_e}\right).
\end{equation}

We present the exclusion limits in the $\lam'_{P}/m_{\tilde{f}}^2=\lam^\prime_{D}/m_{\tilde{f}}^2$ vs. $m_{\tilde
{\chi}^0_1}$ as well as the $\lam^\prime_D/m_{\tilde{f}}^2$ vs. $\lam^\prime_P/m_{\tilde{f}}^2$ plane in 
Fig.~\ref{fig:131-112-model-dep}. As in the previous scenario for the latter plane we consider three neutralino 
masses, which are 1000\,MeV, 3000\,MeV, and 5000\,MeV here. In the $\lam'_{P}/m_{\tilde{f}}^2=\lam^\prime
_{D}/m_{\tilde{f}}^2$ vs. $m_{\tilde{\chi^0_1}}$ plane we observe a similar pattern as before. The sensitivity reach 
is mostly independent of the neutralino mass, except for the region close to the meson masses. The reach is bounded 
from above as the neutralino would decay too fast and below, where the neutralino production and the decay 
would be insufficient. We can extend the sensitivity reach slightly, if we were able to detect neutral final states. 

\texttt{MAPP2} enhances the sensitivity reach of \texttt{MAPP1} due to the increased volume and integrated luminosity. 
However, both forms of the \texttt{MAPP} detector are only sensitive beyond the current single coupling limits by factors 
between 5 and 10, but not beyond the coupling product limit of this scenario for a sfermion mass of $m_{\tilde{f}}=1 \,\mathrm{TeV}$.
This lack of sensitivity of the \texttt{MAPP} detectors to reach the current coupling product limit, which depends on the 
selectron mass only, can potentially be reduced to some extent, if non-degenerate sfermion masses are assumed and the lightest 
allowed values of sfermions masses are taken. Even in that limit, we do not expect the MAPP programs to exceed the coupling 
product limit by much.
\texttt{ANUBIS} has the greatest reach out of all 3 detectors, which extends beyond the coupling limits. 
Next, we consider the $\lam^\prime_D/m_{\tilde{f}}^2$ vs. $\lam^\prime_P/m_{\tilde{f}}^2$ plane. An important difference 
compared to the first scenario is that now the production coupling does not lead to neutralino decay modes and 
consequently we are not bounded from the right side for large values of $\lam^\prime_P$ in 
Fig.~\ref{fig:131-112-model-dep}. Comparing with the current bounds on the RPV couplings, \texttt{ANUBIS} may 
explore parameter regions that are still allowed, while the sensitive regions of \texttt{MAPP1} and \texttt{MAPP2} are 
almost completely ruled out by the current limit on the product of the two RPV couplings for $m_{\tilde{f}}=
1\,$TeV.  For the medium neutralino mass at 3000\,MeV, \texttt{ANUBIS} may probe $\lam'_{131}/m^2_{\tilde{f}}$ ($\lam'_{112}/m^2_{\tilde{f}}$) down to $7 
\times 10^{-11}$ GeV$^{-2}$ ($4 \times 10^{-11}$ GeV$^{-2}$) at the upper limit of $\lam'_{112}/m^2_{\tilde{f}}$ 
($\lam'_{131}/m^2_{\tilde{f}}$) for $m^2_{\tilde{f}}=5$ TeV.

\begin{figure}[t]
	\centering
	\includegraphics[width=.999\linewidth]{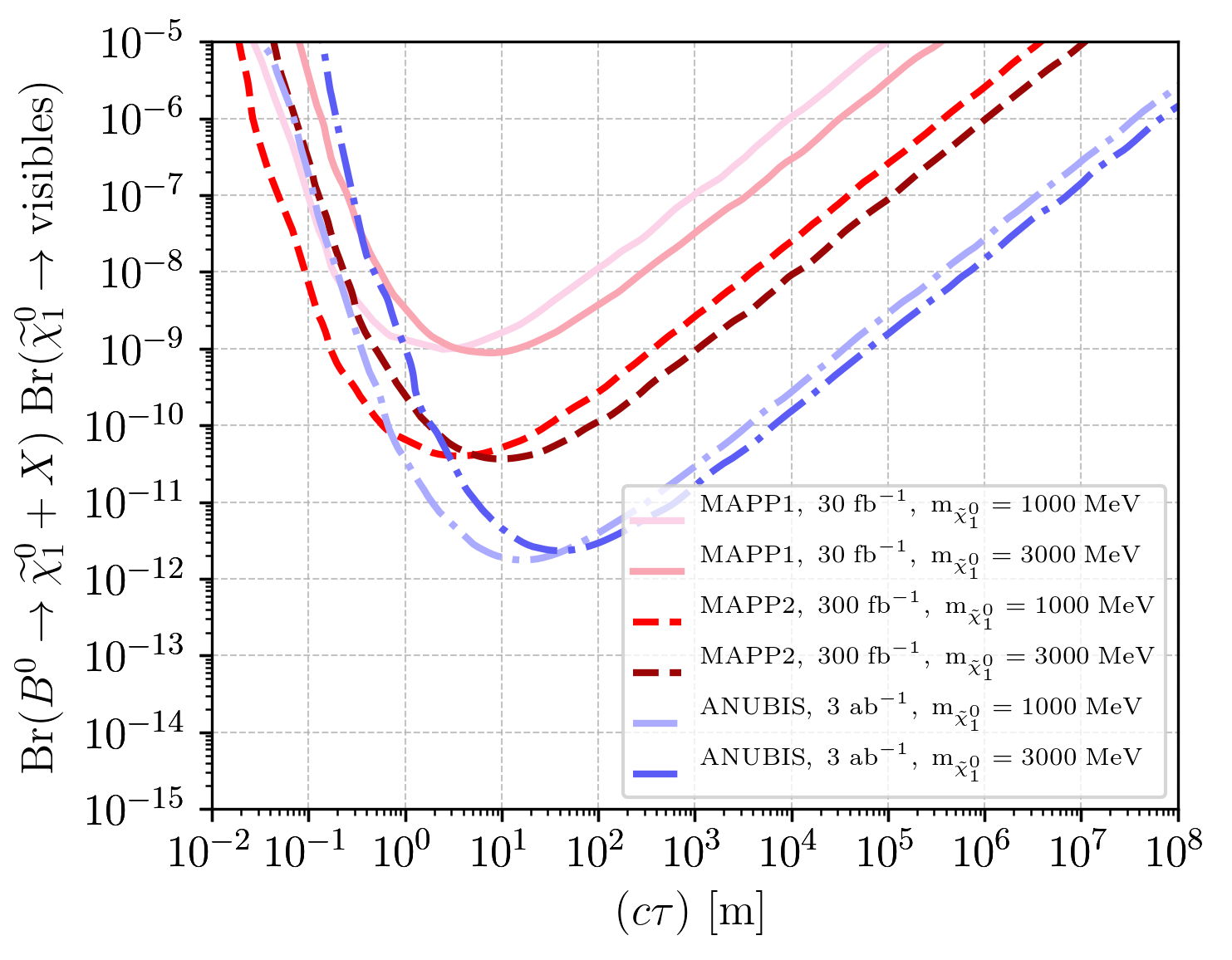}
	\caption{Sensitivity estimate for a neutral long-lived fermion in a model-independent description for the charmed benchmark scenario 2 using our neutralino estimates. Labeling is similar to Fig.~\ref{fig:122-112-model-indep}. For benchmark scenario 2 the considered neutralino masses  are $1$ and $3$ GeV. }\label{fig:131-112-model-indep}
\end{figure}

Lastly, we consider the representation for topologically identical theoretical models in Fig.~\ref{fig:131-112-model-indep} for $m_{\tilde{\chi}_1^0}=$ 1000 MeV and 3000 MeV.
The \texttt{MAPP1} and \texttt{MAPP2} lowest reach differs by more than one order of magnitude, while the 
\texttt{ANUBIS} reach in Br$\,\simeq 2\times 10^{-12}$  is even stronger again by more than another order of magnitude. 
Considering the average lengths to the detector $\Braket{L}$ from Eq.~\eqref{eqn:avg_det_lengths} and average boosts 
$\Braket{\beta\gamma}$ from Tab.~\ref{tab:average_neutralino_boost} for a neutralino mass $m_{\tilde{\chi}^0_1}=1~
\mathrm{GeV}$, the position of the lowest reach should be
\begin{equation}\label{eqn:valley_positions_bottomed}
(c\tau)_{\mathrm{min}}=\left\{  \begin{matrix}
\hspace{2mm} 9.69 \text{ m}&\hspace{3.5mm}\text{ for \texttt{ANUBIS},}\\
\hspace{2mm} 1.45 \text{ m}&\hspace{2mm}\text{ for \texttt{MAPP1},}\\
\hspace{2mm} 1.74 \text{ m}&\hspace{2mm}\text{ for \texttt{MAPP2}.}
\end{matrix}\right.
\end{equation}
This approximately coincides with the valley positions in Fig.~\ref{fig:131-112-model-indep}.

\subsection{Comparison to previously considered Detectors}

In previous works several detectors were studied for the same benchmark scenarios, see Refs.~\cite{Dercks:2018wum,deVries:2015mfw,Dercks:2018eua}. 
Here we  compare those results with \texttt{MAPP1}, \texttt{MAPP2}, and \texttt{ANUBIS} for neutralino masses of 1200 MeV and 3000 MeV in the respective benchmark scenarios. We consider both a model-dependent $\lam^\prime_D/
m_{\tilde{f}}^2$ vs. $\lam^\prime_P/m_{\tilde{f}}^2$ and the model-independent Br vs. $c\tau$ representation. 

\begin{figure}
\centering
  \includegraphics[width=.999\linewidth]{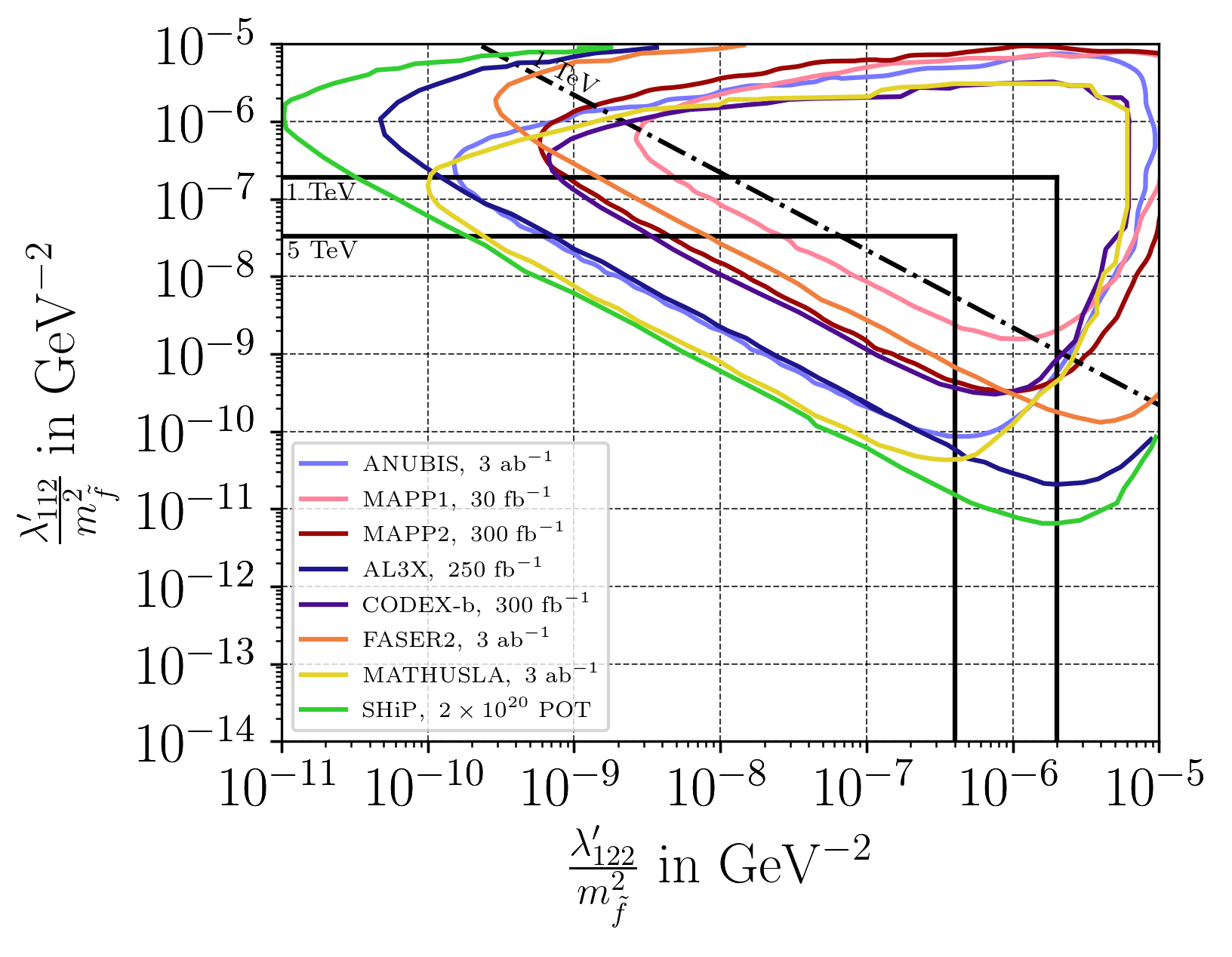}
  \includegraphics[width=.999\linewidth]{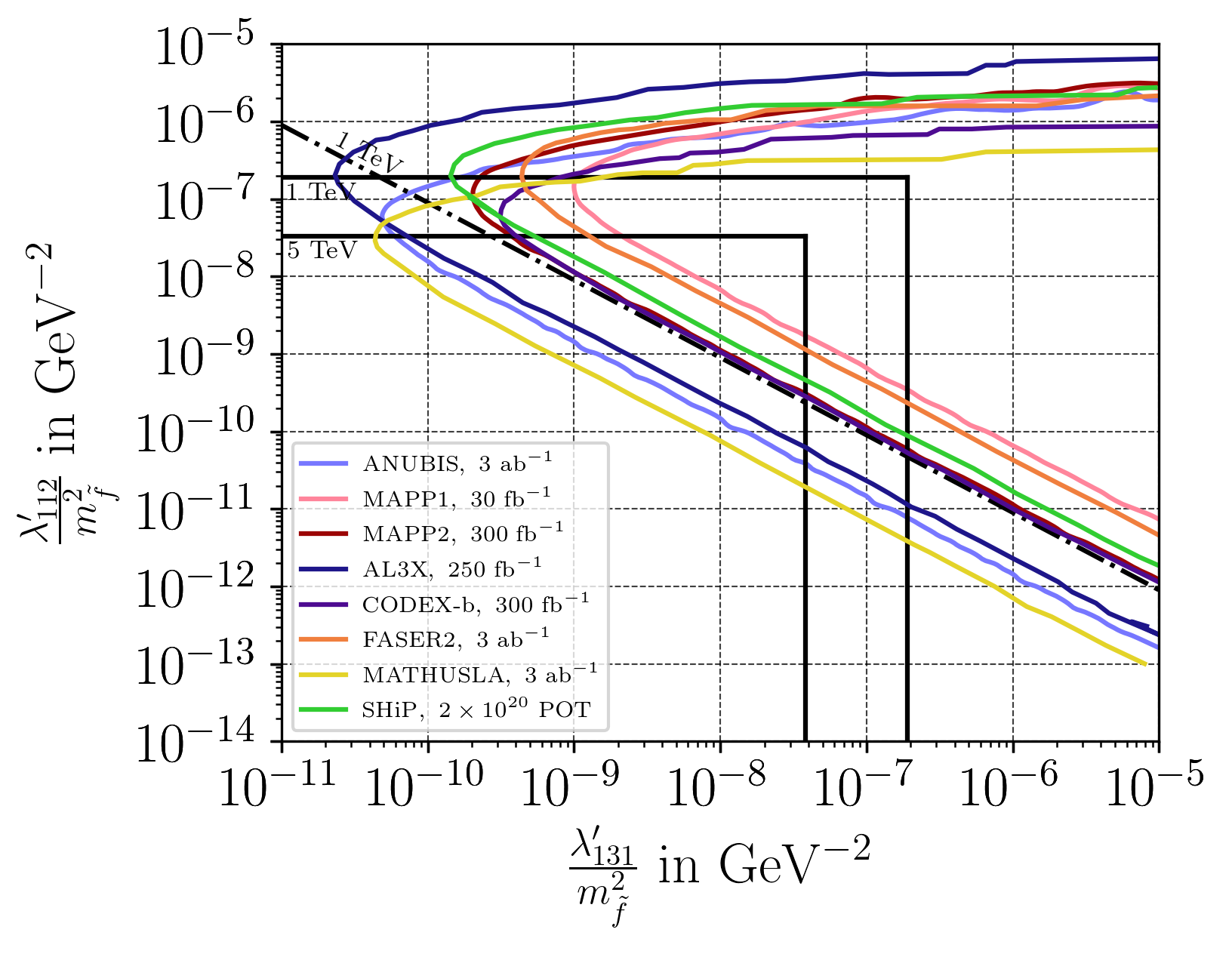}
\caption{Comparison of model-dependent numerical results. The top figure is for benchmark scenario 1, Tab.~\ref{tab:benchmark_scenario_charm} for $m_{\tilde{\chi}_1^0}=1200$ MeV, the lower for benchmark scenario 2, Tab.~\ref{tab:benchmark_scenario_bottom} for $m_{\tilde{\chi}_1^0}=3000$ MeV.} \label{fig:scen_c_and_b_detector_comparison_lvl}
\end{figure}

Firstly, we  look at the model-dependent representation in Fig.~\ref{fig:scen_c_and_b_detector_comparison_lvl}. Benchmark scenario 1, 
Tab.~\ref{tab:benchmark_scenario_charm}, is shown in the top figure, benchmark scenario 2, Tab.~\ref{tab:benchmark_scenario_bottom}, 
in the lower. For both scenarios \texttt{MAPP1} has the lowest reach while \texttt{MAPP2} can substantially extend that reach to parameter 
regions comparable to other detectors, namely \texttt{CODEX-B}, \texttt{FASER2} in the charmed scenario, and \texttt{SHiP} in the bottom 
meson scenario. The advantage of \texttt{MAPP1} is, that the detector is already approved to be implemented for the LHC Run-3. The 
proximity to the \texttt{ATLAS} interaction point combined with the high integrated luminosity of 3\,ab$^{-1}$, however, propels 
\texttt{ANUBIS} to be the most promising proposed detector from this consideration. Only \texttt{AL3X} and \texttt{MATHUSLA} in both 
scenarios, and \texttt{SHiP} for the first scenario, can extend the existing sensitivity reach by a similar amount.

The model-independent representation is shown in Fig.~\ref{fig:scen_c_and_b_detector_comparison_br_ctau}. Again, benchmark scenario 1, 
is shown in the top figure, benchmark scenario 2 in the lower. For the first scenario, summarized in Tab.~\ref{tab:benchmark_scenario_charm}, a hash grid is added in the upper right corner, \textit{i.e.} for large $c\tau$ combined with a large product of branching ratios. 
This region is theoretically excluded and can not be probed in this scenario.
The reason is the following.
In order to continue the curves into the upper right, one must increase both the product of production branching ratios and the decay length. 
If one increases the production coupling $\lam'_P=\lam'_{122}$, then in the charmed scenario this also induces additional invisible decays, \textit{cf.} Tab.~\ref{tab:benchmark_scenario_charm}, which \textit{reduces} the decay length.
This can be compensated by decreasing the decay coupling $\lam'_D=\lam'_{112}$.
But first of all this also reduces the product of production branching. Thus the increase in $\lam'_P$ must be larger than the decrease in $\lam'_D$.
At some point $\lam'_P$ dominates the decay length computation and the reduction in $c\tau$ can no 
longer be compensated by reducing $\lam'_D$.

As before, compared to the other proposed experiments,  
\texttt{MAPP1} has the lowest sensitivity reach for both scenarios because of its smaller angular coverage and lower integrated luminosity. 
However, \texttt{MAPP1} is one of the few proposed far-detector programs that have been approved at CERN.  \texttt{MAPP2} extends the 
sensitivity reach in both directions. \texttt{ANUBIS} shows a similar  behavior to \texttt{MATHUSLA} and is very promising to extend the reach for
detecting any general long-lived particle.

\section{Conclusions}
\label{sect:conclusion}

 \begin{figure}[t]
	\centering
	\includegraphics[width=.999\linewidth]{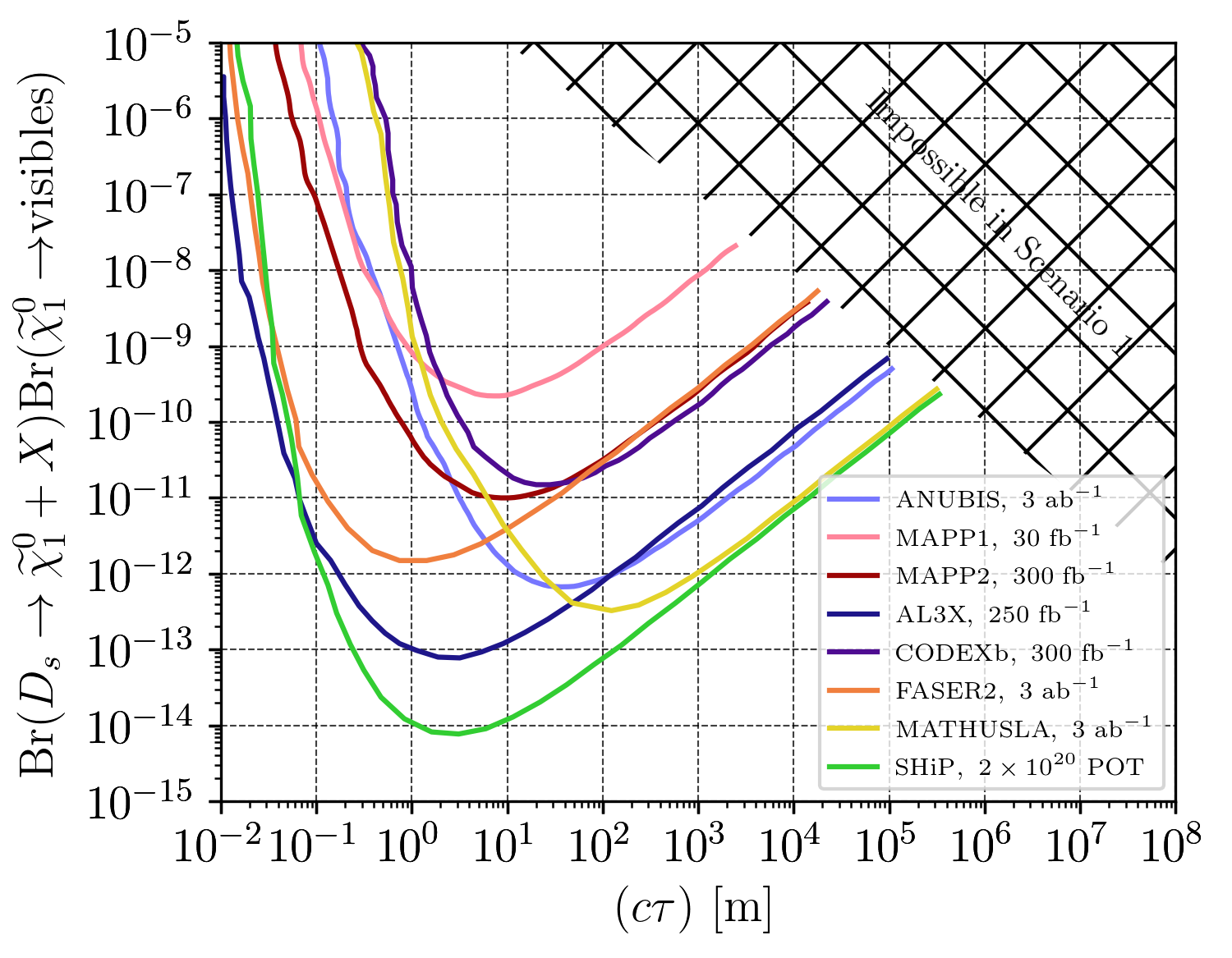}
	\includegraphics[width=.999\linewidth]{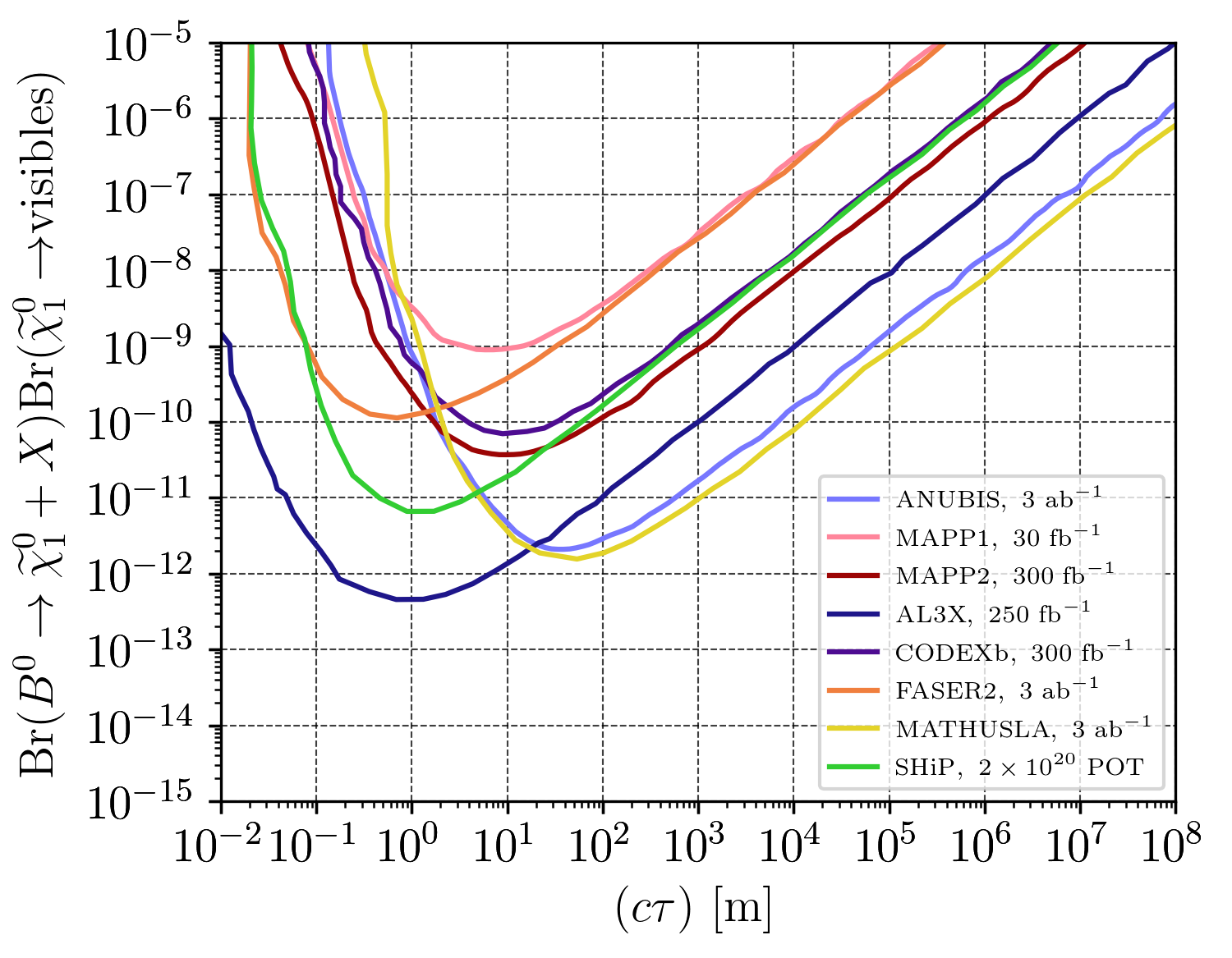}
	\caption{Comparison of Br vs. $c\tau$ numerical results for various proposed detectors. The top figure is for benchmark scenario 1, 
	\textit{cf.} Tab.~\ref{tab:benchmark_scenario_charm}, for $m_{\tilde{\chi}_1^0}=1200$ MeV,
		the lower for benchmark scenario 2, \textit{cf.} Tab.~\ref{tab:benchmark_scenario_bottom}, for $m_{\tilde{\chi}_1^0}=3000$ MeV.
		For the charmed scenario the region for large branching ratio products and large $c\tau$ is theoretically excluded.
		This region is marked by the hashed region.} \label{fig:scen_c_and_b_detector_comparison_br_ctau}
\end{figure}

In this paper we have investigated the potential of the experiments \texttt{ANUBIS}, \texttt{MAPP1}, and \texttt{MAPP2} for the detection
of long-lived light supersymmetric neutralinos produced via rare meson decays. This is an extension of previous works for the same model at other present and proposed experiments: \texttt{ATLAS}, \texttt{SHiP}, \texttt{FASER}, \texttt{CODEX-b}, \texttt{MATHUSLA}, and \texttt{AL3X}
\cite{deVries:2015mfw,Dercks:2018eua,Dercks:2018wum}. The neutralino decays via R-parity violating couplings to a lighter meson and a charged lepton. Following Refs.~\cite{deVries:2015mfw,Dercks:2018eua}, we consider two benchmark scenarios related to either charm or 
bottom mesons decays into the light neutralino.

We find that \texttt{MAPP1} can strengthen current bounds on RPV-couplings in the charmed benchmark scenario. Its planned upgrade program 
\texttt{MAPP2} may extend the sensitivity reach by one order of magnitude in $\lambda'/m^2_{\tilde{f}}$ compared to \texttt{MAPP1}, as a result 
of a greater integrated luminosity and increased solid-angle coverage. The sensitivity range of \texttt{ANUBIS} is shown to be the largest 
among the three detectors in both scenarios.

We compared the exclusion limits of these detectors to those of other experiments derived in Refs.~\cite{Dercks:2018wum,deVries:2015mfw,Dercks:2018eua}.
\texttt{MAPP1} is approved and will explore the parameter space. \texttt{MAPP2} can go beyond this by more than an order of magnitude, and
\texttt{ANUBIS} by yet another order of magnitude in Br, the product of production and decay branching ratios, reaching about $7\times10^{-13}$. But the 
potentially most sensitive experiment here is \texttt{SHiP}, followed by \texttt{AL3X} and \texttt{MATHUSLA}. See Fig.~\ref{fig:scen_c_and_b_detector_comparison_br_ctau}.

In the bottom scenario \texttt{MAPP1} goes down to about $2 \times 10^{-10}$ in Br,  and \texttt{MAPP2} extends this by more than an order of magnitude.
Again \texttt{ANUBIS} can extend this by more than another order of magnitude reaching values as low as Br $\sim3\times 10^{-12}$. Here \texttt{SHiP} suffers
from the lower production of B-mesons and the most sensitive proposed experiment is \texttt{AL3X}, followed by \texttt{MATHUSLA} and almost identically
\texttt{ANUBIS}. In particular, \texttt{AL3X} can achieve this with an integrated luminosity more than one order of magnitude lower than that of \texttt{MATHUSLA} and \texttt{ANUBIS}, see Fig.~\ref{fig:scen_c_and_b_detector_comparison_br_ctau}.

%%%%%%%%%%%%%%%%%%%%%%%%%%%%%%%%%%%%%%%%%%%%%%%%%%%%%%%%%%%%%%%%%%%%%%
\bigskip
\centerline{\bf Acknowledgements}

\bigskip
We thank Vasiliki Mitsou for useful discussions. The work of H.\,K.\,D. was supported through the BMBF-Project 05H18PDCA1.
Z.\,S.\,W. is supported partly by the Ministry of Science and Technology (MoST) of Taiwan with grant number MoST-109-2811-M-007-509, and partly by the Ministry of Science, ICT \& Future Planning of Korea, the Pohang City Government, and the Gyeongsangbuk-do Provincial Government through the Young Scientist Training Asia Pacific Economic Cooperation program of the Asia Pacific Center for Theoretical Physics.

\bigskip
%%%%%%%%%%%%%%%%%%%%%%%%%%%%%%%%%%%%%%%%%%%%%%%%%%%%%%%%%%%%%%%%%%%%%%

%\bibliography{refs}

\begin{thebibliography}{10}

\bibitem{Alimena:2019zri}
J.~Alimena {\em et~al.},
\newblock (2019), arXiv:1903.04497.

\bibitem{Lee:2018pag}
L.~Lee, C.~Ohm, A.~Soffer, and T.-T. Yu,
\newblock Prog. Part. Nucl. Phys. {\bf 106}, 210 (2019), arXiv:1810.12602.

\bibitem{Curtin:2018mvb}
D.~Curtin {\em et~al.},
\newblock Rept. Prog. Phys. {\bf 82}, 116201 (2019), arXiv:1806.07396.

\bibitem{Choudhury:1995pj}
D.~Choudhury and S.~Sarkar,
\newblock Phys. Lett. B {\bf 374}, 87 (1996), arXiv:hep-ph/9511357.

\bibitem{Choudhury:1999tn}
D.~Choudhury, H.~K. Dreiner, P.~Richardson, and S.~Sarkar,
\newblock Phys. Rev. D {\bf 61}, 095009 (2000), arXiv:hep-ph/9911365.

\bibitem{Belanger:2002nr}
G.~Belanger, F.~Boudjema, A.~Pukhov, and S.~Rosier-Lees,
\newblock {A Lower limit on the neutralino mass in the MSSM with nonuniversal
  gaugino masses},
\newblock in {\em {10th International Conference on Supersymmetry and
  Unification of Fundamental Interactions (SUSY02)}}, pp. 919--924, 2002,
  arXiv:hep-ph/0212227.

\bibitem{Hooper:2002nq}
D.~Hooper and T.~Plehn,
\newblock Phys. Lett. B {\bf 562}, 18 (2003), arXiv:hep-ph/0212226.

\bibitem{Bottino:2002ry}
A.~Bottino, N.~Fornengo, and S.~Scopel,
\newblock Phys. Rev. D {\bf 67}, 063519 (2003), arXiv:hep-ph/0212379.

\bibitem{Belanger:2003wb}
G.~Belanger, F.~Boudjema, A.~Cottrant, A.~Pukhov, and S.~Rosier-Lees,
\newblock JHEP {\bf 03}, 012 (2004), arXiv:hep-ph/0310037.

\bibitem{Vasquez:2010ru}
D.~Albornoz~Vasquez, G.~Belanger, C.~Boehm, A.~Pukhov, and J.~Silk,
\newblock Phys. Rev. D {\bf 82}, 115027 (2010), arXiv:1009.4380.

\bibitem{Calibbi:2013poa}
L.~Calibbi, J.~M. Lindert, T.~Ota, and Y.~Takanishi,
\newblock JHEP {\bf 10}, 132 (2013), arXiv:1307.4119.

\bibitem{Gogoladze:2002xp}
I.~Gogoladze, J.~D. Lykken, C.~Macesanu, and S.~Nandi,
\newblock Phys. Rev. D {\bf 68}, 073004 (2003), arXiv:hep-ph/0211391.

\bibitem{Dreiner:2009ic}
H.~K. Dreiner {\em et~al.},
\newblock Eur. Phys. J. C {\bf 62}, 547 (2009), arXiv:0901.3485.

\bibitem{Grifols:1988fw}
J.~Grifols, E.~Masso, and S.~Peris,
\newblock Phys. Lett. B {\bf 220}, 591 (1989).

\bibitem{Ellis:1988aa}
J.~R. Ellis, K.~A. Olive, S.~Sarkar, and D.~Sciama,
\newblock Phys. Lett. B {\bf 215}, 404 (1988).

\bibitem{Lau:1993vf}
K.~Lau,
\newblock Phys. Rev. D {\bf 47}, 1087 (1993).

\bibitem{Dreiner:2003wh}
H.~Dreiner, C.~Hanhart, U.~Langenfeld, and D.~R. Phillips,
\newblock Phys. Rev. D {\bf 68}, 055004 (2003), arXiv:hep-ph/0304289.

\bibitem{Dreiner:2013tja}
H.~K. Dreiner, J.-F. Fortin, J.~Isern, and L.~Ubaldi,
\newblock Phys. Rev. D {\bf 88}, 043517 (2013), arXiv:1303.7232.

\bibitem{Profumo:2008yg}
S.~Profumo,
\newblock Phys. Rev. D {\bf 78}, 023507 (2008), arXiv:0806.2150.

\bibitem{Dreiner:2011fp}
H.~K. Dreiner, M.~Hanussek, J.~S. Kim, and S.~Sarkar,
\newblock Phys. Rev. D {\bf 85}, 065027 (2012), arXiv:1111.5715.

\bibitem{Bechtle:2015nua}
P.~Bechtle {\em et~al.},
\newblock Eur. Phys. J. C {\bf 76}, 96 (2016), arXiv:1508.05951.

\bibitem{Dreiner:1997uz}

\newblock H.~K. Dreiner{\em {An Introduction to explicit R-parity violation}}
  Vol.~21 (, 2010), pp. 565--583, arXiv:hep-ph/9707435.

\bibitem{Barbier:2004ez}
R.~Barbier {\em et~al.},
\newblock Phys. Rept. {\bf 420}, 1 (2005), arXiv:hep-ph/0406039.

\bibitem{Mohapatra:2015fua}

\newblock R.~N. Mohapatra{\em {Supersymmetry and R-parity: an Overview}}
  Vol.~90 (, 2015), p. 088004, arXiv:1503.06478.

\bibitem{Dedes:2001zia}
A.~Dedes, H.~K. Dreiner, and P.~Richardson,
\newblock Phys. Rev. D {\bf 65}, 015001 (2001), arXiv:hep-ph/0106199.

\bibitem{Dreiner:2002xg}
H.~Dreiner, G.~Polesello, and M.~Thormeier,
\newblock (2002), arXiv:hep-ph/0207160.

\bibitem{Dreiner:2009er}
H.~Dreiner {\em et~al.},
\newblock Phys. Rev. D {\bf 80}, 035018 (2009), arXiv:0905.2051.

\bibitem{Alekhin:2015byh}
S.~Alekhin {\em et~al.},
\newblock Rept. Prog. Phys. {\bf 79}, 124201 (2016), arXiv:1504.04855.

\bibitem{Gorbunov:2015mba}
D.~Gorbunov and I.~Timiryasov,
\newblock Phys. Rev. D {\bf 92}, 075015 (2015), arXiv:1508.01780.

\bibitem{deVries:2015mfw}
J.~de~Vries, H.~K. Dreiner, and D.~Schmeier,
\newblock Phys. Rev. D {\bf 94}, 035006 (2016), arXiv:1511.07436.

\bibitem{Gligorov:2017nwh}
V.~V. Gligorov, S.~Knapen, M.~Papucci, and D.~J. Robinson,
\newblock Phys. Rev. D {\bf 97}, 015023 (2018), arXiv:1708.09395.

\bibitem{Aielli:2019ivi}
G.~Aielli {\em et~al.},
\newblock (2019), arXiv:1911.00481.

\bibitem{Feng:2017uoz}
J.~L. Feng, I.~Galon, F.~Kling, and S.~Trojanowski,
\newblock Phys. Rev. D {\bf 97}, 035001 (2018), arXiv:1708.09389.

\bibitem{Ariga:2018uku}
FASER, A.~Ariga {\em et~al.},
\newblock Phys. Rev. D {\bf 99}, 095011 (2019), arXiv:1811.12522.

\bibitem{Ariga:2019ufm}
FASER, A.~Ariga {\em et~al.},
\newblock (2019), arXiv:1901.04468.

\bibitem{Chou:2016lxi}
J.~P. Chou, D.~Curtin, and H.~Lubatti,
\newblock Phys. Lett. B {\bf 767}, 29 (2017), arXiv:1606.06298.

\bibitem{Gligorov:2018vkc}
V.~V. Gligorov, S.~Knapen, B.~Nachman, M.~Papucci, and D.~J. Robinson,
\newblock Phys. Rev. D {\bf 99}, 015023 (2019), arXiv:1810.03636.

\bibitem{Helo:2018qej}
J.~C. Helo, M.~Hirsch, and Z.~S. Wang,
\newblock JHEP {\bf 07}, 056 (2018), arXiv:1803.02212.

\bibitem{Dercks:2018eua}
D.~Dercks, J.~De~Vries, H.~K. Dreiner, and Z.~S. Wang,
\newblock Phys. Rev. D {\bf 99}, 055039 (2019), arXiv:1810.03617.

\bibitem{Dercks:2018wum}
D.~Dercks, H.~K. Dreiner, M.~Hirsch, and Z.~S. Wang,
\newblock Phys. Rev. D {\bf 99}, 055020 (2019), arXiv:1811.01995.

\bibitem{Wang:2019orr}
Z.~S. Wang and K.~Wang,
\newblock Phys. Rev. D {\bf 101}, 115018 (2020), arXiv:1904.10661.

\bibitem{Wang:2019xvx}
Z.~S. Wang and K.~Wang,
\newblock Phys. Rev. D {\bf 101}, 075046 (2020), arXiv:1911.06576.

\bibitem{Sirunyan:2019gut}
CMS, A.~M. Sirunyan {\em et~al.},
\newblock Phys. Lett. B {\bf 797}, 134876 (2019), arXiv:1906.06441.

\bibitem{Aad:2019xav}
ATLAS, G.~Aad {\em et~al.},
\newblock Phys. Rev. D {\bf 101}, 052013 (2020), arXiv:1911.12575.

\bibitem{Bauer:2019vqk}
M.~Bauer, O.~Brandt, L.~Lee, and C.~Ohm,
\newblock (2019), arXiv:1909.13022.

\bibitem{Hirsch:2020klk}
M.~Hirsch and Z.~S. Wang,
\newblock Phys. Rev. D {\bf 101}, 055034 (2020), arXiv:2001.04750.

\bibitem{Staelens:2019gzt}
MoEDAL, M.~Staelens,
\newblock {Recent Results and Future Plans of the MoEDAL Experiment},
\newblock in {\em {Meeting of the Division of Particles and Fields of the
  American Physical Society}}, 2019, arXiv:1910.05772.

\bibitem{Bartl:1988cn}
A.~Bartl, W.~Majerotto, and N.~Oshimo,
\newblock Phys. Lett. B {\bf 216}, 233 (1989).

\bibitem{Allanach:2003eb}
B.~Allanach, A.~Dedes, and H.~Dreiner,
\newblock Phys. Rev. D {\bf 69}, 115002 (2004), arXiv:hep-ph/0309196,
\newblock [Erratum: Phys.Rev.D 72, 079902 (2005)].

\bibitem{Ibanez:1991pr}
L.~E. Ibanez and G.~G. Ross,
\newblock Nucl. Phys. B {\bf 368}, 3 (1992).

\bibitem{Dreiner:2012ae}
H.~K. Dreiner, M.~Hanussek, and C.~Luhn,
\newblock Phys. Rev. D {\bf 86}, 055012 (2012), arXiv:1206.6305.

\bibitem{Dreiner:2008ca}
H.~K. Dreiner and S.~Grab,
\newblock Phys. Lett. B {\bf 679}, 45 (2009), arXiv:0811.0200.

\bibitem{Dercks:2017lfq}
D.~Dercks, H.~Dreiner, M.~E. Krauss, T.~Opferkuch, and A.~Reinert,
\newblock Eur. Phys. J. C {\bf 77}, 856 (2017), arXiv:1706.09418.

\bibitem{Allanach:1999mh}
B.~Allanach, A.~Dedes, and H.~K. Dreiner,
\newblock Phys. Rev. D {\bf 60}, 056002 (1999), arXiv:hep-ph/9902251,
\newblock [Erratum: Phys.Rev.D 86, 039906 (2012)].

\bibitem{Allanach:1999ic}
B.~Allanach, A.~Dedes, and H.~K. Dreiner,
\newblock Phys. Rev. D {\bf 60}, 075014 (1999), arXiv:hep-ph/9906209.

\bibitem{Barger:1989rk}
V.~D. Barger, G.~Giudice, and T.~Han,
\newblock Phys. Rev. D {\bf 40}, 2987 (1989).

\bibitem{Bhattacharyya:1997vv}
G.~Bhattacharyya,
\newblock {A Brief review of R-parity violating couplings},
\newblock in {\em {Workshop on Physics Beyond the Standard Model: Beyond the
  Desert: Accelerator and Nonaccelerator Approaches}}, pp. 194--201, 1997,
  arXiv:hep-ph/9709395.

\bibitem{Kao:2009fg}
Y.~Kao and T.~Takeuchi,
\newblock (2009), arXiv:0910.4980.

\bibitem{Domingo:2018qfg}
F.~Domingo {\em et~al.},
\newblock JHEP {\bf 02}, 066 (2019), arXiv:1810.08228.

\bibitem{Bansal:2019zak}
S.~Bansal, A.~Delgado, C.~Kolda, and M.~Quiros,
\newblock Phys. Rev. D {\bf 100}, 093005 (2019), arXiv:1906.01063.

%\cite{ATLAS:2015gma}
\bibitem{ATLAS:2015gma}
[ATLAS],
%``Constraints on promptly decaying supersymmetric particles with lepton-number- and R-parity-violating interactions using Run-1 ATLAS data,''
ATLAS-CONF-2015-018.
%15 citations counted in INSPIRE as of 02 Feb 2021

%\cite{Khachatryan:2016ycy}
\bibitem{Khachatryan:2016ycy}
V.~Khachatryan \textit{et al.} [CMS],
%``Search for R-parity violating decays of a top squark in proton-proton collisions at $\sqrt{s}$ = 8 TeV,''
Phys. Lett. B \textbf{760} (2016), 178-201
doi:10.1016/j.physletb.2016.06.039
[arXiv:1602.04334 [hep-ex]].
%8 citations counted in INSPIRE as of 02 Feb 2021

\bibitem{Aaij:2015bpa}
LHCb, R.~Aaij {\em et~al.},
\newblock JHEP {\bf 03}, 159 (2016), arXiv:1510.01707,
\newblock [Erratum: JHEP 09, 013 (2016), Erratum: JHEP 05, 074 (2017)].

\bibitem{Aaij:2016avz}
LHCb, R.~Aaij {\em et~al.},
\newblock Phys. Rev. Lett. {\bf 118}, 052002 (2017), arXiv:1612.05140,
\newblock [Erratum: Phys.Rev.Lett. 119, 169901 (2017)].

\bibitem{Cacciari:1998it}
M.~Cacciari, M.~Greco, and P.~Nason,
\newblock JHEP {\bf 05}, 007 (1998), arXiv:hep-ph/9803400.

\bibitem{Cacciari:2001td}
M.~Cacciari, S.~Frixione, and P.~Nason,
\newblock JHEP {\bf 03}, 006 (2001), arXiv:hep-ph/0102134.

\bibitem{Cacciari:2012ny}
M.~Cacciari {\em et~al.},
\newblock JHEP {\bf 10}, 137 (2012), arXiv:1205.6344.

\bibitem{Cacciari:2015fta}
M.~Cacciari, M.~L. Mangano, and P.~Nason,
\newblock Eur. Phys. J. C {\bf 75}, 610 (2015), arXiv:1507.06197.

\bibitem{Sjostrand:2007gs}
T.~Sjostrand, S.~Mrenna, and P.~Z. Skands,
\newblock Comput. Phys. Commun. {\bf 178}, 852 (2008), arXiv:0710.3820.

\bibitem{Sjostrand:2006za}
T.~Sjostrand, S.~Mrenna, and P.~Z. Skands,
\newblock JHEP {\bf 05}, 026 (2006), arXiv:hep-ph/0603175.

\end{thebibliography}
\bibliographystyle{h-physrev5}

\end{document}